%% file: main.tex
\pdfoutput=1
\RequirePackage{fix-cm}
\documentclass[fleqn,usenatbib]{mnras}
\usepackage[T1]{fontenc}
\usepackage{amsmath}
\usepackage{newtxtext,newtxmath}
\usepackage{graphicx,tikz}
\makeatletter
\def\endfigure{\end@float}
\def\endtable{\end@float}
\def\@printed{Preprint \tod@y}
\makeatother
\usepackage[caption=false]{subfig}
\usepackage{placeins}

\title[Centrifugally accelerated particles on dipolar trajectories]{Relativistic Dynamics of Centrifugally Accelerated Particles Along Rotating Dipolar Trajectories}
\author[E. Gaprindashvili et al.]{
Elene Gaprindashvili,$^{1,2,3}$
Natia Kevlishvili,$^{1}$
Giorgi Khomeriki$^{1,4,5,6}$
and Andria Rogava$^{1}$
\\
$^{1}$Institute for Theoretical Physics, Ilia State University, G. Tsereteli 3, 0162 Tbilisi, Georgia\\
$^{2}$Ivane Javakhishvili Tbilisi State University, 3 Chavchavadze Ave., 0179 Tbilisi, Georgia\\
$^{3}$School of Physics, Free University of Tbilisi, 240 David Agmashenebeli Alley, 0159 Tbilisi, Georgia\\
$^{4}$E. Kharadze Georgian National Astrophysical Observatory, Abastumani 0301, Georgia\\
$^{5}$Department of Physics, Kansas State University, 1228 N. Martin Luther King Jr. Drive, Manhattan,\\ KS 66506-2601, USA\\
$^{6}$Vekua School of Physics and Mathematics, 9 Tchaikovsky St., 0105 Tbilisi, Georgia
}
\date{Draft}
\pubyear{2026}
\begin{document}
\label{firstpage}
\pagerange{\pageref{firstpage}--\pageref{lastpage}}
\maketitle

\begin{abstract}
We study a bead that slides without friction along a field line of a rigidly rotating dipole, as an idealized model of a particle in a pulsar magnetosphere. The centrifugal force pushes the bead away from the rotation axis. Wherever the line reaches a local maximum of its distance from that axis inside the light cylinder, the bead can therefore oscillate about that point. For arbitrary magnetic inclination and azimuth the equilibrium angles are the roots of a cubic; we give their stability and the range of lines on which they lie between the star and the light cylinder. At the cylinder the outcome depends on how the line bends: on a line that leads the rotation the bead approaches the cylinder only asymptotically, whereas on a trailing line it crosses in a finite time at subluminal speed. Wells become more common as the magnetic axis tilts towards orthogonality, but they rarely lie between a stellar footpoint and the first crossing of the cylinder, and a bead whose Jacobi energy exceeds its rest energy passes all of them. A simple score for emission from the outer magnetosphere is largest at inclinations between about $60^\circ$ and $75^\circ$, depending on the assumed yield; this range mainly reflects how many of the sampled lines reach the cylinder. Published radio and gamma-ray inclinations for the same pulsars differ by more than the effect we look for, so present samples cannot test it.
\end{abstract}

\input{01_introduction}
\input{02a_general_treatment}
\input{02_model}
\input{04_equilibria}
\input{05_special_cases}
\input{03_trajectory_results}
\input{06_numerical_results}
\input{08_tilt_dependence}
\input{07_equilibrium_geometry}
\input{08b_observed_inclinations}
\input{09_conclusions}

\section*{Acknowledgements}
This work was supported by the Shota Rustaveli National Science Foundation of Georgia (SRNSFG) [grant number FR-24-2858].

\bibliographystyle{mnras}
\bibliography{references}
\appendix
\input{appendices}
\label{lastpage}
\end{document}

%% file: 01_introduction.tex
\section{Introduction}
\label{sec:introduction}

Pulsars are powered by rotation \citep{Pacini1967,Gold1968}. The magnetic field of the neutron star couples that rotation to the surrounding plasma and drives currents and an outflow \citep{GoldreichJulian1969,Michel1973}. Close to the star the field is nearly dipolar and the plasma corotates with it. This cannot continue out to the light cylinder, where corotation would require the speed of light. Even in vacuum the field of an inclined rotator is retarded there \citep{Deutsch1955}. In plasma-filled solutions the outer field lines open into a wind with an equatorial current sheet, and both the wind and the spin-down torque depend on the magnetic inclination \citep{Contopoulos1999,Spitkovsky2006,Tchekhovskoy2013}. Kinetic models add that the supply of pairs affects the currents and the sites of particle acceleration \citep{Philippov2015,PhilippovSpitkovsky2018}. A rigidly rotating dipole is therefore not a model of the outer magnetosphere. We use it as a reference geometry in which the effect of rotation on a guided particle can be worked out exactly.

The simplest problem of this kind is a bead on a straight rotating wire. \citet{MachabeliRogava1994} found that the radial acceleration of the bead can reverse sign in flat spacetime, and \citet{KhomerikiRogava2020} obtained exact solutions for a broader set of initial conditions. This flat-space effect is distinct from the reversal of the centrifugal force in curved spacetime \citep{AbramowiczPrasanna1990}. The shape of the wire matters: on a suitable trailing spiral the bead can cross the light cylinder \citep{Rogava2003}. Rotating guides have been applied to particle acceleration in active galactic nuclei \citep{RiegerMannheim2000,OsmanovRogavaBodo2007} and to pulsar radio emission \citep{Gangadhara1996,ThomasGangadhara2007}. They have also motivated centrifugal acceleration models for the Crab and Vela pulsars \citep{OsmanovRieger2009,OsmanovRieger2017,OsmanovRieger2019}, whose pulsed emission extends to very high energies \citep{Aliu2011,Abdalla2018,HESS2023}. For a review of the rotating-wire problem and its astrophysical applications, see \citet{Rogava2023}.

On a dipolar line the situation is different. A line that leaves the star, moves away from the rotation axis and then curves back has a point of greatest distance from the axis, and the centrifugal force pushes a bead towards that point from both sides. Earlier work on straight and spiral guides showed that radiation reaction can produce stable equilibria for charged particles \citep{Dalakishvili2007}. For the dipolar geometry the closest precedent is \citet{Bakhtadze2019}, who found surfaces of stable equilibria in a model with radiative drag, building on their work on rotating guides \citep{Bakhtadze2017}. Drag lets a particle settle; here we ask what happens without it, when the energy in the rotating frame is conserved: a particle released inside a well then oscillates indefinitely, and one that arrives with enough energy passes through.

We consider a massive bead on a rigidly rotating dipolar line of arbitrary magnetic inclination and azimuth. Section~\ref{sec:general} first treats the general problem: a bead on a line of arbitrary shape and rotation law in a stationary, axisymmetric spacetime. It derives the equation of motion, the condition for equilibrium and, for rigid rotation, the Jacobi integral and the stability of the equilibria. Section~\ref{sec:model} specializes these results to a dipolar line in flat spacetime. Sections~\ref{sec:general_equilibria} and \ref{sec:special_cases} locate the wells of the effective potential and give their stability. Section~\ref{sec:bead_trajectory} follows individual trajectories, including the approach to the light cylinder and the passage through it. Sections~\ref{sec:numerical_equilibria} and \ref{sec:tilt_concentration} map the wells and count them as a function of inclination, both on whole lines and on the path from a stellar footpoint to the first crossing of the cylinder. The model is deliberately idealized, and Section~\ref{sec:discussion} states what it leaves out before any comparison with data. Section~\ref{sec:observed_inclinations} then asks whether published inclination measurements could reveal a geometric effect of this kind, and Section~\ref{sec:conclusions} summarizes.

%% file: 02a_general_treatment.tex
\section{General treatment}
\label{sec:general}

Before turning to the rotating dipole, we formulate the problem of a bead on a rotating line in a general setting. A treatment of this kind was given by \citet{KhomerikiRogava2020}. They embedded a prescribed trajectory $\vartheta(r)$, $\phi(r)+\Omega(r)t$ in a stationary, axisymmetric metric, derived the induced metric and, for rigid rotation, the conserved energy, and then solved the straight rotating wire in flat spacetime exactly. Here we present the formalism in a more compact and transparent form. We write the equation of motion in closed form for an arbitrary parametrization of the line and an arbitrary rotation law, and we also derive the conditions for equilibrium and stability, which were not considered there.

The results of this section hold for an arbitrary shape of the line, an arbitrary rotation law along it and any background spacetime of the class defined below. The flat-space dipole studied in the rest of the paper is then obtained by specializing three ingredients: the metric, the rotation law and the shape of the line.

\subsection{Background spacetime and the rotating line}

We consider a stationary, axisymmetric spacetime with coordinates $(t,r,\vartheta,\phi)$, where $t$ is the time of a distant observer and $\phi$ the azimuth about the rotation axis. We assume that the line element has the form
\begin{equation}
ds^2=g_{tt}\,dt^2+2g_{t\phi}\,dt\,d\phi+g_{\phi\phi}\,d\phi^2+g_{rr}\,dr^2+g_{\vartheta\vartheta}\,d\vartheta^2,
\label{eq:general_background}
\end{equation}
with components that depend on $r$ and $\vartheta$ only. We use the signature $(-,+,+,+)$ and units with $c=1$. The only off-diagonal component is $g_{t\phi}$, so the class contains Minkowski spacetime in spherical coordinates, the Schwarzschild metric, the Kerr metric in Boyer--Lindquist coordinates and the metrics of slowly rotating stars. To avoid confusion with the magnetic colatitude $\theta$ used from Section~\ref{sec:model} on, we denote the polar angle of the background coordinates by $\vartheta$.

Points of the line are labelled by a parameter $\lambda$. At time $t$ the line occupies
\begin{equation}
r=\tilde r(\lambda),\qquad
\vartheta=\tilde\vartheta(\lambda),\qquad
\phi=\tilde\phi(\lambda)+\Omega(\lambda)\,t .
\label{eq:general_line}
\end{equation}
The functions $\tilde r$, $\tilde\vartheta$ and $\tilde\phi$ give the shape of the line at $t=0$, and $\Omega(\lambda)$ is the angular velocity of the point $\lambda$ about the rotation axis. We allow $\Omega$ to vary along the line, which describes differential rotation; rigid rotation is the special case $\Omega'=0$. Here and below a prime denotes $d/d\lambda$.

Any parameter that increases monotonically along the line may serve as $\lambda$. The most direct choice is the radial coordinate itself, $\lambda=r$, so that $\tilde r(\lambda)=\lambda$ and the line is given by $\tilde\vartheta(r)$, $\tilde\phi(r)$ and $\Omega(r)$. This choice is natural for lines that run monotonically outwards, such as open field lines or spirals. It fails wherever $r$ stops changing along the line, $\tilde r'=0$: there $r$ cannot label the points of the line, and $dr/dt$ vanishes whatever the motion of the bead. A dipolar line $r=R\sin^2\theta$ has such a point at the magnetic equator, which is exactly where the equilibria of Section~\ref{sec:general_equilibria} lie. We therefore keep $\lambda$ general here, use $\lambda=r$ only in the aligned examples of Appendices~\ref{app:aligned_flat} and \ref{app:aligned_schwarzschild}, and take $\lambda$ to be the magnetic colatitude $\theta$ from Section~\ref{sec:model} on.

\subsection{Induced metric}

Along the line, Eq.~\eqref{eq:general_line} gives
\begin{equation}
dr=\tilde r'\,d\lambda,\qquad
d\vartheta=\tilde\vartheta'\,d\lambda,\qquad
d\phi=\zeta\,d\lambda+\Omega\,dt,
\end{equation}
where
\begin{equation}
\zeta(\lambda,t)\equiv\tilde\phi'+\Omega'\,t .
\end{equation}
Substituting into Eq.~\eqref{eq:general_background}, the line element on the world sheet swept out by the line becomes
\begin{equation}
ds^2=G_{tt}\,dt^2+2G_{t\lambda}\,dt\,d\lambda+G_{\lambda\lambda}\,d\lambda^2,
\label{eq:general_induced}
\end{equation}
with
\begin{align}
G_{tt}&=g_{tt}+2\Omega\,g_{t\phi}+\Omega^2g_{\phi\phi},\label{eq:general_Gtt}\\
G_{t\lambda}&=\zeta\left(g_{t\phi}+\Omega\,g_{\phi\phi}\right),\label{eq:general_Gtl}\\
G_{\lambda\lambda}&=g_{rr}\,\tilde r'^2+g_{\vartheta\vartheta}\,\tilde\vartheta'^2+g_{\phi\phi}\,\zeta^2.\label{eq:general_Gll}
\end{align}
The background components are evaluated on the line, at $\bigl(\tilde r(\lambda),\tilde\vartheta(\lambda)\bigr)$. For $\lambda=r$ one has $\tilde r'=1$.

$G_{tt}$ is the squared norm of the four-velocity direction $\partial_t+\Omega\,\partial_\phi$ of an observer who corotates with the point $\lambda$. It vanishes where corotation would require the speed of light, which in flat spacetime is the light cylinder. $G_{t\lambda}$ measures how far the line is tilted in azimuth in the corotating frame, and $G_{\lambda\lambda}\,d\lambda^2$ is the interval along the line at fixed $t$.

When $\Omega'\neq0$, the azimuthal pitch $\zeta$ grows linearly with $t$: points further out rotate at a different rate, and the line winds up about the axis without limit. $G_{t\lambda}$ and $G_{\lambda\lambda}$ then depend explicitly on time, whereas $G_{tt}$ does not. For rigid rotation the induced metric is stationary.

Following the notation used in the rest of the paper, we write
\begin{equation}
\begin{gathered}
D\equiv-G_{tt},\qquad B\equiv G_{t\lambda},\qquad A\equiv G_{\lambda\lambda},\\
K\equiv AD+B^2=-\det G_{ab}.
\end{gathered}
\label{eq:general_ABDK}
\end{equation}
The induced metric is Lorentzian, so that the bead can move along the line on a timelike worldline, only where $K>0$.

\subsection{Equations of motion}

The line acts on the bead as an ideal, frictionless constraint: the force it exerts is orthogonal to the world sheet of the line. The worldline of the bead therefore extremizes its proper time among all curves that lie on the world sheet, which means that it is a geodesic of the induced metric~\eqref{eq:general_induced},
\begin{equation}
\frac{d^2x^a}{d\tau^2}+\Gamma^a_{bc}\frac{dx^b}{d\tau}\frac{dx^c}{d\tau}=0,\qquad a,b,c\in\{t,\lambda\},
\label{eq:general_geodesic}
\end{equation}
where
\begin{equation}
\Gamma^a_{bc}=\tfrac12G^{ad}\left(\partial_bG_{dc}+\partial_cG_{db}-\partial_dG_{bc}\right)
\end{equation}
are the Christoffel symbols of $G_{ab}$. For differential rotation they include the time derivatives of $G_{t\lambda}$ and $G_{\lambda\lambda}$.

We define
\begin{equation}
\gamma\equiv u^t=\frac{dt}{d\tau},\qquad v\equiv\frac{d\lambda}{dt},\qquad u^\lambda=\frac{d\lambda}{d\tau}=\gamma v .
\end{equation}
For $\lambda=r$, $v$ is the radial coordinate velocity. In general it is the rate of change of a coordinate, not a physical speed. The normalization $G_{ab}u^au^b=-1$ gives
\begin{equation}
\gamma=\left(-G_{tt}-2G_{t\lambda}v-G_{\lambda\lambda}v^2\right)^{-1/2}
=\left(D-2Bv-Av^2\right)^{-1/2},
\label{eq:general_gamma}
\end{equation}
which requires $D-2Bv-Av^2>0$.

The $t$-component of Eq.~\eqref{eq:general_geodesic} reads
\begin{equation}
\frac{d^2t}{d\tau^2}=-\gamma^2\left(\Gamma^t_{tt}+2\Gamma^t_{t\lambda}v+\Gamma^t_{\lambda\lambda}v^2\right).
\label{eq:general_t_component}
\end{equation}
To obtain the $\lambda$-component in laboratory time, we differentiate $d\lambda/d\tau=\gamma v$:
\begin{equation}
\frac{d^2\lambda}{d\tau^2}=\gamma^2\frac{dv}{dt}+v\,\frac{d^2t}{d\tau^2}.
\end{equation}
Inserting this and Eq.~\eqref{eq:general_t_component} into the $\lambda$-component, $d^2\lambda/d\tau^2=-\gamma^2(\Gamma^\lambda_{tt}+2\Gamma^\lambda_{t\lambda}v+\Gamma^\lambda_{\lambda\lambda}v^2)$, and dividing by $\gamma^2$, we obtain
\begin{equation}
\frac{dv}{dt}=-\Gamma^\lambda_{tt}
+\left(\Gamma^t_{tt}-2\Gamma^\lambda_{t\lambda}\right)v
+\left(2\Gamma^t_{t\lambda}-\Gamma^\lambda_{\lambda\lambda}\right)v^2
+\Gamma^t_{\lambda\lambda}\,v^3 .
\label{eq:general_master}
\end{equation}
This closed equation for $\lambda(t)$ is the central result of this section. It holds for any metric of the form~\eqref{eq:general_background}, any shape of the line and any rotation law $\Omega(\lambda)$; for differential rotation its coefficients depend explicitly on $t$. The $t$-component adds no further information, because geodesic motion preserves the normalization: once $\lambda(t)$ is known, Eq.~\eqref{eq:general_gamma} gives $\gamma$ and hence the proper time.

\subsection{Equilibria}

A bead can remain at a fixed point $\lambda_0$ of the line for all time if $v=0$ and $dv/dt=0$ there, that is, if $\Gamma^\lambda_{tt}=0$. Since $G_{tt}$ does not depend on $t$,
\begin{equation}
\Gamma^\lambda_{tt}=\tfrac12G^{\lambda\lambda}\left(2\,\partial_tG_{t\lambda}-\partial_\lambda G_{tt}\right),
\qquad G^{\lambda\lambda}=\frac{D}{K}.
\end{equation}
A bead at rest on the line must lie inside the light surface, $D>0$, so $G^{\lambda\lambda}\neq0$ and the equilibrium condition is
\begin{equation}
\partial_\lambda G_{tt}=2\,\partial_tG_{t\lambda}=2\,\Omega'\left(g_{t\phi}+\Omega\,g_{\phi\phi}\right).
\label{eq:general_equilibrium}
\end{equation}
The right-hand side does not depend on $t$. The equilibrium points therefore stay fixed on the line even while a differentially rotating line winds up. For rigid rotation the condition reduces to $D'=0$: the bead can rest only at an extremum of $-G_{tt}$ along the line.

In flat spacetime $g_{tt}=-1$, $g_{t\phi}=0$ and $g_{\phi\phi}=\rho^2$, where $\rho=r\sin\vartheta$ is the distance from the rotation axis. Then $G_{tt}=-1+\Omega^2\rho^2$, and Eq.~\eqref{eq:general_equilibrium} becomes $2\Omega\Omega'\rho^2+2\Omega^2\rho\rho'=2\Omega\Omega'\rho^2$, that is $\rho'=0$. In flat spacetime a bead can rest only where the distance of the line from the rotation axis is stationary, and differential rotation does not move these points. Gravity and frame dragging shift them through $g_{tt}$ and $g_{t\phi}$.

\subsection{Rigid rotation: Jacobi integral and stability}

For rigid rotation, $\Omega'=0$, the induced metric does not depend on $t$, and $t$ is a cyclic coordinate of the Lagrangian $L=\frac12G_{ab}\dot x^a\dot x^b$, where a dot denotes $d/d\tau$. The conjugate momentum gives the conserved quantity
\begin{equation}
\mathcal E\equiv-\frac{\partial L}{\partial\dot t}=D\dot t-B\dot\lambda .
\label{eq:general_jacobi}
\end{equation}
Solving for $\dot t$ and inserting the result into the normalization yields
\begin{equation}
\mathcal E^2=D+K\dot\lambda^{\,2}.
\label{eq:general_energy}
\end{equation}
Where $K>0$, the bead can therefore reach only points with $D\le\mathcal E^2$, and $\sqrt D$ acts as an effective potential. Its turning points lie where $D=\mathcal E^2$, and its extrema are the equilibria found above. For differential rotation no such integral exists, because the line winds up and continually changes the energy of the bead.

The stability of an equilibrium follows from the same integral. Differentiating Eq.~\eqref{eq:general_energy} with respect to $\tau$ and dividing by $\dot\lambda$ gives
\begin{equation}
\ddot\lambda=-\frac{D'+K'\dot\lambda^{\,2}}{2K}.
\end{equation}
Near an equilibrium $\lambda_*$, where $D'(\lambda_*)=0$, a small displacement $\delta\lambda$ therefore obeys
\begin{equation}
\frac{d^2\delta\lambda}{d\tau^2}+\frac{D''(\lambda_*)}{2K(\lambda_*)}\,\delta\lambda=0 .
\label{eq:general_stability}
\end{equation}
The equilibrium is stable if $D''(\lambda_*)>0$, a minimum of the effective potential, and unstable if $D''(\lambda_*)<0$. A bead at rest on the line has $d\tau/dt=\sqrt{D}$, so the frequency of small oscillations in laboratory time is
\begin{equation}
\omega_t^{\,2}=\frac{D(\lambda_*)\,D''(\lambda_*)}{2K(\lambda_*)}.
\label{eq:general_frequency}
\end{equation}
The position of the equilibrium depends on $D$ alone, but the frequency also depends on $A$ and $B$ through $K$. For differential rotation the linearized equation has coefficients that change with time as the line winds up, and its stability is not decided by the sign of a single quantity.

For a stationary induced metric the Christoffel symbols can be written in terms of $A$, $B$ and $D$. The coefficients of Eq.~\eqref{eq:general_master} become
\begin{align}
-\Gamma^\lambda_{tt}&=-\frac{DD'}{2K},&
\Gamma^t_{tt}-2\Gamma^\lambda_{t\lambda}&=\frac{3BD'}{2K},\nonumber\\
2\Gamma^t_{t\lambda}-\Gamma^\lambda_{\lambda\lambda}&=\frac{2AD'-DA'-2BB'}{2K},&
\Gamma^t_{\lambda\lambda}&=\frac{BA'-2AB'}{2K}.
\label{eq:general_rigid_coefficients}
\end{align}

\subsection{From the general treatment to the rotating dipole}

The rest of the paper specializes this framework in three steps. First, we take Minkowski spacetime, for which $D=1-\Omega^2\rho^2$ and the light surface $D=0$ is the light cylinder $\rho=1/\Omega$. Second, we take rigid rotation, so that the Jacobi integral~\eqref{eq:general_energy} exists. Third, we take a dipolar field line and use the magnetic colatitude as the parameter, $\lambda=\theta$. Because the magnetic axis is inclined, the shape functions $\tilde r(\theta)$, $\tilde\vartheta(\theta)$ and $\tilde\phi(\theta)$ are not written out explicitly; they follow from the tilted Cartesian representation of Section~\ref{sec:model}. With these choices, $A$, $B$ and $D$ of Section~\ref{sec:model} are the components~\eqref{eq:general_ABDK} of the induced metric. Equation~\eqref{eq:general_energy} becomes the first integral~\eqref{eq:equilibrium_energy_integral}, Eq.~\eqref{eq:general_master} with the coefficients~\eqref{eq:general_rigid_coefficients} becomes Eq.~\eqref{eq:general_theta_lab}, The equilibrium condition becomes $C'=0$, which is the starting point of Section~\ref{sec:general_equilibria}, and the stability criterion~\eqref{eq:general_stability} becomes the condition $C''<0$ used there.

%% file: 02_model.tex
\section{The rotating dipole in flat spacetime}
\label{sec:model}
\subsection{Physical assumptions}

We now apply the general treatment of Section~\ref{sec:general} to the case studied in the rest of the paper: flat spacetime, rigid rotation and a dipolar line parametrized by the magnetic colatitude. A massive bead slides without friction along a dipolar field line that rotates at constant angular velocity $\Omega>0$ about the laboratory $z$-axis. The magnetic axis makes an angle $\alpha$ with the rotation axis. We work in Minkowski spacetime, set $c=1$ and take the rest mass of the bead to be unity. The only force on the bead is the constraint that keeps it on the line. We leave out gravity, any electric field along the line and radiation reaction, and return to these omissions in Section~\ref{sec:discussion}. The shape of the line is prescribed for all time, so the calculation isolates the effect of geometry, as in constrained-motion models for other guides and backgrounds \citep{Rogava2003,ArsenadzeOsmanov2017}.

A line is labelled by its spherical radius $R$ at the magnetic equator and by its magnetic azimuth $\phi_m$. Position along the line is given by the magnetic colatitude $\theta$. Where numbers are quoted, lengths are in units of the light-cylinder radius $\rho_{\rm LC}=1/\Omega$.

\subsection{Geometry of the field line}

In the magnetic frame, a dipolar field line obeys
\begin{equation}
r_m=R\sin^2\theta.
\end{equation}
Its Cartesian coordinates are
\begin{align}
x_m &= R\sin^3\theta\cos\phi_m,\\
y_m &= R\sin^3\theta\sin\phi_m,\\
z_m &= R\sin^2\theta\cos\theta.
\end{align}

We tilt the magnetic frame by $\alpha$ and then rotate it through
the angle $\Omega t$. Define
\begin{align}
F &= x_m\cos\alpha+z_m\sin\alpha,\\
G &= y_m,\\
H &= -x_m\sin\alpha+z_m\cos\alpha.
\end{align}
The bead's laboratory position is
\begin{align}
x &= F\cos(\Omega t)-G\sin(\Omega t),\\
y &= F\sin(\Omega t)+G\cos(\Omega t),\\
z &= H.
\end{align}

The perpendicular distance from the rotation axis is
\begin{equation}
\rho=\sqrt{x^2+y^2}=\sqrt{F^2+G^2}.
\end{equation}

\subsection{Induced metric and Lagrangian}

Starting with the Minkowski line element
\begin{equation}
ds^2=-dt^2+dx^2+dy^2+dz^2,
\end{equation}
we substitute the prescribed position of the bead. The result is
\begin{equation}
ds^2=-D(\theta)\,dt^2
     +2B(\theta)\,dt\,d\theta
     +A(\theta)\,d\theta^2,
\end{equation}
where
\begin{align}
A &= F'^2+G'^2+H'^2
   =R^2\sin^2\theta(1+3\cos^2\theta),\\
B &= \Omega(FG'-GF')
   =\Omega R^2\sin\alpha\sin\phi_m\sin^4\theta,\\
C &= \Omega^2(F^2+G^2)=\Omega^2\rho^2,\\
D &= 1-C.
\end{align}
A prime denotes differentiation with respect to $\theta$. This is the induced metric~\eqref{eq:general_induced} with $\lambda=\theta$, $D=-G_{tt}$, $B=G_{t\lambda}$ and $A=G_{\lambda\lambda}$. In particular $D=1-\Omega^2\rho^2$ is the flat-space form of Eq.~\eqref{eq:general_Gtt}, and $A=|d\boldsymbol r/d\theta|^2$ is the squared arc length of the line per unit $\theta$ in the corotating frame.

The mixed term has a simple meaning: $B=\Omega\rho^2\,d\varphi_{\rm rot}/d\theta$, where $\varphi_{\rm rot}$ is the azimuth of the line about the rotation axis in the corotating frame. Where $B>0$ the line advances in the sense of rotation as $\theta$ increases; where $B<0$ it falls behind. Lines in the tilt plane, $\phi_m=0$, have $B=0$.

Using proper time $\tau$ as the parameter, the bead's
quadratic Lagrangian is
\begin{equation}
L=\frac{1}{2}
\left(-D\dot{t}^{\,2}
      +2B\dot{t}\dot{\theta}
      +A\dot{\theta}^{\,2}\right).
\end{equation}
A dot denotes differentiation with respect to $\tau$.

\subsection{Equations of motion}

The Euler--Lagrange equation for $\theta$ gives
\begin{equation}
A\ddot{\theta}+B\ddot{t}
+\frac{1}{2}A'\dot{\theta}^{\,2}
+\frac{1}{2}D'\dot{t}^{\,2}=0.
\end{equation}

Because the Lagrangian does not explicitly depend on $t$, the
quantity
\begin{equation}
\mathcal E=D\dot{t}-B\dot{\theta}
\end{equation}
is the Jacobi integral per unit rest mass: $\mathcal E=E_{\rm lab}-\Omega L_z$, where $E_{\rm lab}=\gamma$ and $L_z=\gamma\rho v_\varphi$ are the inertial energy and axial angular momentum per unit mass. The rotating guide can exchange both with the bead, with $dE_{\rm lab}=\Omega\,dL_z$.

For a massive bead, proper-time normalization requires
\begin{equation}
-D\dot{t}^{\,2}
+2B\dot{t}\dot{\theta}
+A\dot{\theta}^{\,2}=-1.
\end{equation}
Combining this condition with the conserved $\mathcal E$ yields
\begin{equation}
\mathcal E^2=D+(AD+B^2)\dot{\theta}^{\,2}.\label{eq:equilibrium_energy_integral}
\end{equation}
This is the general first integral~\eqref{eq:general_energy} with $\lambda=\theta$. It is exact, and most of what follows is read from it.

\subsubsection{Explicit proper-time equations}
Differentiating the conserved energy supplies the second equation
needed to solve for both accelerations:
\begin{equation}
D\ddot t-B\ddot\theta
=B'\dot\theta^{\,2}+C'\dot\theta\dot t.
\end{equation}
Define $K=AD+B^2$. Solving the two coupled equations gives
\begin{align}
\ddot\theta={}&\frac{1}{2K}\Bigl[
DC'\dot t^{\,2}-2BC'\dot\theta\dot t
\nonumber\\
&\hspace{22mm}-(DA'+2BB')\dot\theta^{\,2}\Bigr],
\label{eq:general_theta_proper}\\
\ddot t={}&\frac{1}{2K}\Bigl[
BC'\dot t^{\,2}+2AC'\dot\theta\dot t
\nonumber\\
&\hspace{22mm}+(2AB'-BA')\dot\theta^{\,2}\Bigr].
\label{eq:general_time_proper}
\end{align}
Together with the normalization, these equations describe the
motion for arbitrary fixed $\alpha$, $\phi_m$, $R$ and $\Omega$.
The laboratory trajectory then follows from $x(\theta,t)$,
$y(\theta,t)$ and $z(\theta)$ given above.

\subsubsection{Explicit laboratory-time equation}
Let $v=d\theta/dt$. This is the rate of change of a coordinate; the
physical speed of the bead is given below. Since
\begin{equation}
\frac{d^2\theta}{dt^2}
=\frac{\ddot\theta-v\ddot t}{\dot t^{\,2}},
\end{equation}
Eqs.~\eqref{eq:general_theta_proper} and
\eqref{eq:general_time_proper} give the closed equation
\begin{equation}
\begin{split}
\frac{d^2\theta}{dt^2}=\frac{1}{2K}\Bigl\{
&DC'-3BC'v\\
&-(2AC'+DA'+2BB')v^2\\
&+(BA'-2AB')v^3\Bigr\}.
\end{split}
\label{eq:general_theta_lab}
\end{equation}
This is Eq.~\eqref{eq:general_master} with the rigid-rotation coefficients~\eqref{eq:general_rigid_coefficients}, written with $D'=-C'$. Appendix~\ref{app:coeffs} gives the coefficients explicitly for the dipole.

For initial data $\theta(t_0)=\theta_0$ and $v(t_0)=v_0$, a
future-directed massive trajectory requires
\begin{equation}
D(\theta_0)-2B(\theta_0)v_0-A(\theta_0)v_0^2>0.
\label{eq:general_eom_initial_condition}
\end{equation}
The relation between laboratory and proper time, and the total
laboratory speed, are
\begin{align}
\frac{d\tau}{dt}&=\sqrt{D-2Bv-Av^2},\\
\left|\frac{d\boldsymbol r}{dt}\right|^2&=C+2Bv+Av^2<1.
\end{align}
Integrating Eq.~\eqref{eq:general_theta_lab} also gives
$\tau(t)$ and the laboratory velocity. The equation of motion is
regular as long as the induced metric is Lorentzian, $K>0$. This
holds at the light cylinder itself when $B\ne0$, because there $D=0$
and $K=B^2>0$; the cylinder is a singular point of the equation only
on lines with $B=0$. Setting $v=0$ in
Eq.~\eqref{eq:general_theta_lab} shows that a bead can stay at a
fixed point of the line only where $C'=0$. This is the flat-space, rigid-rotation form of Eq.~\eqref{eq:general_equilibrium}, and it is the condition examined in the next section.

%% file: 04_equilibria.tex
\section{Equilibrium points on a general rotating dipolar line}
\label{sec:general_equilibria}

In the rotating frame, the centrifugal force pushes the bead away from the rotation axis. Follow a line to a point where its cylindrical radius is locally largest: moving a short distance in either direction takes the bead closer to the axis, so the force along the line points back. A bead displaced a little oscillates about that point; settling there would require dissipation. 

\subsection{Effective potential and equilibrium condition}

In the first integral, Eq.~\eqref{eq:equilibrium_energy_integral}, the term $(AD+B^2)\dot\theta^{\,2}$ is positive whenever the bead moves along the line. Hence $\mathcal E^2\geq D$ everywhere on a trajectory, with equality at a turning point, where $\dot\theta=0$. The quantity
\begin{equation}
V_{\mathrm{eff}}(\theta)
=\sqrt{D(\theta)}
=\sqrt{1-\Omega^2\rho^2(\theta)}
\label{eq:general_equilibrium_potential}
\end{equation}
therefore acts as an effective potential: the bead moves where $V_{\mathrm{eff}}\leq\mathcal E$ and turns back where the two are equal.

Inside the light cylinder $0<D\leq1$, so the potential never exceeds unity, the rest energy of the bead. A bead with $\mathcal E>1$ has no turning point inside the cylinder and cannot be confined. The potential reaches $\mathcal E=1$ only on the rotation axis, $\rho=0$; on a line that crosses the axis, a bead with exactly this energy lies on a separatrix. For an oscillation the bead needs a minimum of the potential with a turning point on each side, both outside the star. The condition $\mathcal E<1$ is thus necessary for confinement but not sufficient.

A bead can stay at a fixed position $\theta=\theta_*$ only at a stationary point of
this potential. Since $\Omega$ is constant and $D(\theta_*)>0$,
\begin{equation}
V_{\mathrm{eff}}'(\theta_*)=0
\quad\Longleftrightarrow\quad
C'(\theta_*)=0
\quad\Longleftrightarrow\quad
\left.\frac{d\rho^2}{d\theta}\right|_{\theta_*}=0.
\label{eq:general_equilibrium_condition}
\end{equation}
Setting $\dot\theta=\ddot\theta=0$ in the equation of motion gives the same condition. The mixed coefficient $B$ affects the motion near an equilibrium but not its position.

A stationary point is a physical equilibrium only if it lies outside the star and inside the light cylinder,
\begin{equation}
R\sin^2\theta_*>R_*,
\qquad
\Omega\rho(\theta_*)<1,
\label{eq:equilibrium_admissibility}
\end{equation}
where $R_*$ is the stellar radius. The second inequality is needed because a bead held at a fixed point of the line moves at the corotation speed $\Omega\rho$. We call a stationary point that satisfies both conditions \emph{allowed}.

\subsection{The equilibrium angles}

Introduce the abbreviations
\begin{equation}
\begin{aligned}
a&=\sin\alpha, & b&=\cos\alpha\cos\phi_m,\\
e&=b^2+\sin^2\phi_m, & m&=ab.
\end{aligned}
\label{eq:general_equilibrium_abem}
\end{equation}
Writing $s=\sin\theta$ and $c_\theta=\cos\theta$, the squared distance
from the rotation axis becomes
\begin{equation}
\rho^2
=R^2s^4
\left(a^2c_\theta^2+2msc_\theta+es^2\right).
\label{eq:rho_squared_polynomial}
\end{equation}
Its derivative gives
\begin{equation}
C'(\theta)=2\Omega^2R^2s^3
\left[
2a^2c_\theta^3+5msc_\theta^2+(3e-a^2)s^2c_\theta-ms^3
\right].
\label{eq:C_prime_before_cubic}
\end{equation}

Away from the equator, $c_\theta\ne0$, we set $u=\tan\theta=s/c_\theta$.
Dividing the square bracket in
Eq.~\eqref{eq:C_prime_before_cubic} by $c_\theta^3$ yields the cubic
polynomial
\begin{equation}
P(u)
=-mu^3+(3e-a^2)u^2+5mu+2a^2.
\label{eq:arbitrary_phi_P}
\end{equation}
The stationary angles away from the equator are therefore the roots of
\begin{equation}
P(u)=0,\qquad u=\tan\theta.
\label{eq:general_equilibrium_cubic}
\end{equation}
Each real root gives one angle in $0<\theta<\pi$: a positive root lies
in the northern half of the line and a negative root in the southern
half.

The tangent substitution misses the equator. Directly from
Eq.~\eqref{eq:C_prime_before_cubic}, $\theta=\pi/2$ is stationary
if and only if
\begin{equation}
m=\sin\alpha\cos\alpha\cos\phi_m=0.
\label{eq:equatorial_stationary_condition}
\end{equation}
The endpoints $\theta=0,\pi$ are excluded: on the ideal line
$r_m=R\sin^2\theta$ both lie at the origin, inside the star.

The stationary angles depend on $\alpha$ and $\phi_m$ only. Changing
$R$ or $\Omega$ rescales the distance of each point from the rotation
axis and so decides whether it satisfies
Eq.~\eqref{eq:equilibrium_admissibility}.

\subsection{Stability}
\label{sec:arbitrary_phi_stability}

At a stationary point, differentiation of
Eq.~\eqref{eq:general_equilibrium_potential} gives
\begin{equation}
V_{\mathrm{eff}}''(\theta_*)
=-\frac{C''(\theta_*)}{2\sqrt{D(\theta_*)}}.
\end{equation}
Therefore, for an equilibrium inside the light cylinder,
\begin{align}
C''(\theta_*)<0
&\quad\Longrightarrow\quad \text{stable},\\
C''(\theta_*)>0
&\quad\Longrightarrow\quad \text{unstable}.
\end{align}
A local maximum of $\rho$ is a minimum of the effective potential and
is stable, as anticipated at the start of this section; a local minimum
of $\rho$ is unstable. If $C''(\theta_*)=0$ the test is inconclusive.

For the roots of the cubic, the sign of $C''$ can be read from the
polynomial itself. Equation~\eqref{eq:C_prime_before_cubic} can be
written as
\begin{equation}
C'(\theta)
=2\Omega^2R^2\sin^3\theta\cos^3\theta\,
P(\tan\theta).
\label{eq:arbitrary_phi_C_prime}
\end{equation}
At a root $u_i=\tan\theta_i$ of $P$,
\begin{align}
C''(\theta_i)
&=2\Omega^2R^2\sin^3\theta_i\cos^3\theta_i
  (1+u_i^2)P'(u_i),\\
P'(u)
&=-3mu^2+2(3e-a^2)u+5m.
\label{eq:arbitrary_phi_C_second}
\end{align}
Hence
\begin{align}
\cos\theta_i\,P'(u_i)<0
&\quad\Longrightarrow\quad \text{stable},\\
\cos\theta_i\,P'(u_i)>0
&\quad\Longrightarrow\quad \text{unstable}.
\label{eq:arbitrary_phi_root_stability}
\end{align}

Small displacements $\delta\theta$ from an equilibrium obey, in proper
time,
\begin{equation}
\frac{d^2\delta\theta}{d\tau^2}
-\frac{C''(\theta_*)}{2K_*}\delta\theta=0,
\label{eq:general_equilibrium_linear_motion}
\end{equation}
which is Eq.~\eqref{eq:general_stability} with $\lambda=\theta$ and $D''=-C''$. Here
\begin{equation}
D_*=D(\theta_*),\qquad
K_*=A(\theta_*)D_*+B(\theta_*)^2.
\end{equation}
A bead at rest on the line has $d\tau/dt=\sqrt{D_*}$, so about a stable
equilibrium the frequency of small oscillations in laboratory time is
\begin{equation}
\omega_{\mathrm{small},t}^{\,2}
=-\frac{D_*C''(\theta_*)}{2K_*}.
\label{eq:general_equilibrium_lab_frequency}
\end{equation}
The mixed coefficient $B$ enters this frequency through $K_*$, although
it does not affect where the equilibrium lies.

\subsubsection{Number of stationary points.}
For $m\ne0$, let $q=3e-a^2$. The discriminant of the cubic is
\begin{equation}
\Delta
=
500m^4
+m^2\left(25q^2-180a^2q-108a^4\right)
-8a^2q^3.
\label{eq:arbitrary_phi_discriminant}
\end{equation}
For a tilted line with $0<\alpha<\pi/2$ and $m\ne0$, the sign of
$\Delta$ decides the count. If $\Delta>0$ there are three stationary
points, two stable and one unstable. If $\Delta<0$ there is one, and it
is stable. At $\Delta=0$ two of them merge at a point where
$C'=C''=0$. These counts refer to the complete ideal line; the stellar
surface and the light cylinder can remove some of the points.

The unstable point is where the line passes closest to the rotation axis. The line actually crosses the axis, $F=G=0$, only if $\sin\phi_m=0$, that is, in the tilt plane. Off that plane the closest approach is at $\rho>0$, and it is unstable for the same reason: the centrifugal force pushes the bead away from it on both sides. As $\phi_m$ moves away from the tilt plane, this minimum of $\rho$ and a neighbouring maximum approach each other and disappear together at $\Delta=0$; Section~\ref{sec:numerical_equilibria} gives an example.

\subsubsection{Allowed values of $R$.}
The stationary angle $\theta_i$ fixes the ratio $h_i=\rho_i/R$:
\begin{equation}
h_i=
\sin^2\theta_i
\sqrt{
\left(a\cos\theta_i+b\sin\theta_i\right)^2
+\sin^2\theta_i\sin^2\phi_m
}.
\label{eq:arbitrary_phi_h}
\end{equation}
For $h_i>0$ the point lies outside the star and inside the light
cylinder when
\begin{equation}
\frac{R_*}{\sin^2\theta_i}
<R<
\frac{1}{\Omega h_i}.
\label{eq:arbitrary_phi_R_interval}
\end{equation}
If $h_i=0$ there is no upper bound. As $R$ grows from small values, the point first emerges from the star and later leaves through the light cylinder; if the interval is empty, the point is never allowed. Conversely, the point can be placed at any chosen fraction $0<f<1$ of the light-cylinder radius, $\rho_i=f\rho_{\rm LC}$, by taking $R=f/(\Omega h_i)$, provided the lower bound is respected. A bead held there has the corotation Lorentz factor $\gamma_{\rm cor}=(1-f^2)^{-1/2}$.

\subsubsection{Equatorial special case.}
When $m=0$, the equator must be added to the roots of $P(u)$. At
$\theta=\pi/2$,
\begin{equation}
\rho_{\mathrm{eq}}=R\sqrt e,
\qquad
C''\left(\frac{\pi}{2}\right)
=2\Omega^2R^2(a^2-3e).
\end{equation}
The equatorial point is stable if $a^2<3e$ and unstable if
$a^2>3e$; at equality, higher-order terms decide. It is allowed when
$R>R_*$ and $\Omega R\sqrt e<1$.

%% file: 05_special_cases.tex
\section{Special cases and limits}
\label{sec:special_cases}

The aligned rotator, the orthogonal rotator and lines in the tilt plane can be solved in closed form, and we use them as checks on the numerical maps below. Appendix~\ref{app:straight} recovers the straight rotating wire of \citet{MachabeliRogava1994} as a limit of the orthogonal case.

\subsection{Aligned rotator}

For $\alpha=0$, the magnetic and rotation axes coincide. Then
\begin{align}
B &= 0,\\
C &= \Omega^2R^2\sin^6\theta,
\end{align}
and hence
\begin{equation}
V_{\mathrm{eff}}(\theta)
=\sqrt{1-\Omega^2R^2\sin^6\theta}.
\end{equation}

The distance from the rotation axis grows steadily from each pole to
the magnetic equator, so the only equilibrium is at $\theta=\pi/2$.
It is stable, and it is allowed when the point is outside the star
and $\Omega R<1$.

\subsection{Orthogonal rotator}

For $\alpha=\pi/2$, the magnetic axis lies in the plane
perpendicular to the rotation axis. The coefficients become
\begin{align}
A &= R^2\sin^2\theta(1+3\cos^2\theta),\\
B &= \Omega R^2\sin^4\theta\sin\phi_m,\\
C &= \Omega^2R^2\sin^4\theta
\left(\cos^2\theta+
\sin^2\theta\sin^2\phi_m\right).
\end{align}

For the field line in the tilt plane, $\phi_m=0$, we have
$B=0$ and
\begin{equation}
V_{\mathrm{eff}}(\theta)
=\sqrt{1-\Omega^2R^2\sin^4\theta\cos^2\theta}.
\end{equation}

This line has three stationary points:
\begin{equation}
\theta=\arctan\sqrt{2},\qquad
\theta=\frac{\pi}{2},\qquad
\theta=\pi-\arctan\sqrt{2}.
\end{equation}
The magnetic equator, where the line crosses the rotation axis, is
unstable. The two points on either side of it are stable and have
\begin{equation}
\rho=\frac{2R}{3\sqrt{3}}.
\end{equation}
They are inside the light cylinder if
\begin{equation}
\Omega R<\frac{3\sqrt{3}}{2}.
\end{equation}

\subsection{Field lines in the tilt plane}

For $\phi_m=0$, the field line lies in the plane containing the
magnetic and rotation axes. In this case $G=B=0$, and
\begin{equation}
\rho^2
=R^2\sin^4\theta\sin^2(\theta+\alpha).
\label{eq:tilt_plane_rho}
\end{equation}
For $0<\alpha<\pi/2$, one stationary point is
\begin{equation}
\theta=\pi-\alpha.
\end{equation}
There $\rho=0$: the line crosses the rotation axis. The point is
unstable, and it is an equilibrium only if it lies outside the star.

The other two stationary angles satisfy
\begin{equation}
\tan^2\theta
-3\cot\alpha\,\tan\theta-2=0,
\end{equation}
and therefore
\begin{equation}
\tan\theta=
\frac{3\cot\alpha\pm\sqrt{9\cot^2\alpha+8}}{2}.
\label{eq:tilt_plane_two_roots}
\end{equation}
The positive root lies in the northern half of the line and the
negative root in the southern half. Both are local maxima of $\rho$
and hence stable wherever they are allowed.

For $\phi_m=180^\circ$, the line lies in the same plane on the
opposite side of the magnetic axis, and it crosses the rotation axis
at $\theta=\alpha$.

%% file: 03_trajectory_results.tex
\section{Trajectories and the light cylinder}
\label{sec:bead_trajectory}

\subsection{Initial conditions}
We integrate the laboratory-time equation, Eq.~\eqref{eq:general_theta_lab}, for a bead released at rest on the line with
\begin{equation}
\begin{gathered}
\alpha=45^\circ,\quad \phi_m=100^\circ,\quad
R=2\rho_{\rm LC},\quad R_*=0.01\rho_{\rm LC},\\
\theta_0=10^\circ,\qquad V_{\parallel,0}=0,
\end{gathered}
\end{equation}
with $\Omega=1$ and $t_0=\tau_0=0$.
An intermediate tilt avoids the special symmetries of the aligned and orthogonal cases. The line is long enough to reach the light cylinder, and at this azimuth its outgoing part bends forwards, in the sense of rotation ($B>0$). Reflecting the azimuth in the tilt plane, $\phi_m\to360^\circ-\phi_m$, reverses the bend and leaves the potential unchanged, which lets us isolate the effect of the bend in Section~\ref{sec:motion_regimes}. The stellar radius corresponds to a fast pulsar: for a star of radius 10 km, $R_*/\rho_{\rm LC}=0.01$ at a period of about 20 ms, and the ratio is smaller for slower pulsars.

Here $V_{\parallel,0}$ is the signed speed along the line measured
by a local corotating observer at the launch point. It fixes the
conserved energy and the initial coordinate rate through
\begin{align}
\mathcal E&=\frac{\sqrt{D_0}}{\sqrt{1-V_{\parallel,0}^2}},\\
\left.\frac{d\theta}{dt}\right|_0
&=\frac{V_{\parallel,0}D_0}
{\sqrt{K_0}+V_{\parallel,0}B_0}.
\end{align}
A bead at rest on the rotating line still moves in the laboratory, at the corotation speed $\Omega\rho_0$. The launch point is close to the rotation axis, so this speed is a few per cent of the speed of light and $\mathcal E$ is only slightly below unity.

The integration covers $0\leq\Omega t\leq10$. As a check, quadrature of the first integral gives the same trajectory over this interval.

\subsection{Spatial motion and the two clocks}
\begin{figure*}
\centering
\includegraphics[width=\textwidth]{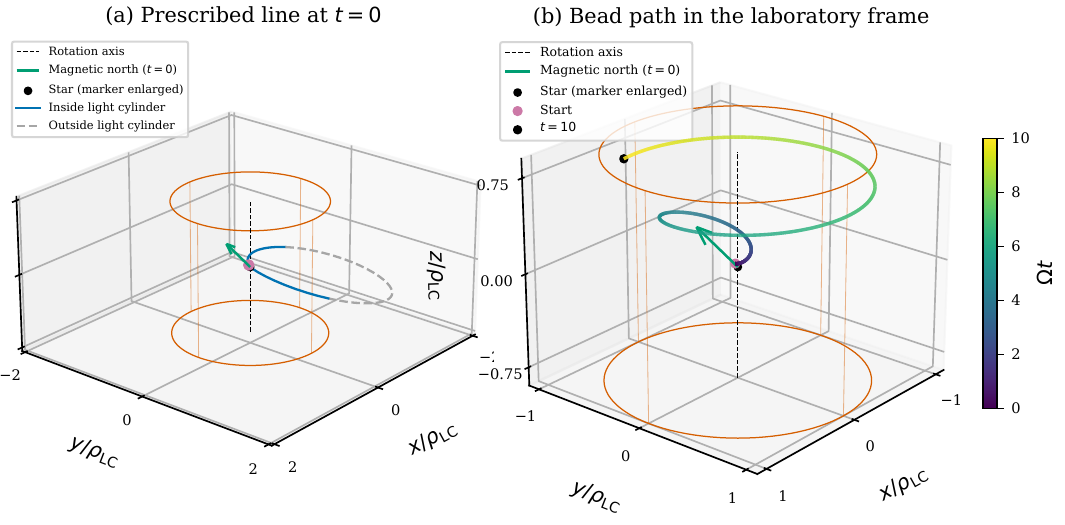}
\caption{The field line and the path of the bead for the initial data in Section~\ref{sec:bead_trajectory}. Left: the dipolar line at $t=0$, solid inside the light cylinder and dashed outside. Right: the path of the bead over $0\leq\Omega t\leq10$, coloured by laboratory time, from the magenta point to the black point. Orange curves outline the light cylinder, the dashed line is the rotation axis and the arrow is the magnetic axis at $t=0$. Lengths are in units of $\rho_{\rm LC}$; the star is drawn enlarged, and the two panels have different scales.}
\label{fig:bead_path}
\end{figure*}

Figure~\ref{fig:bead_path} shows the line at $t=0$ and the path of the
bead in the laboratory. The two are different curves, because the
line rotates while the bead slides along it. The bead first moves
away from the star, towards larger $\theta$ and $\rho$. Near the
light cylinder its progress along the line, $d\theta/dt$, slows down
while the rotation still carries it around the axis. The path winds onto a circle on the light cylinder, at the height where the line meets it.
The magnetic-axis arrow in the figure is drawn for $t=0$; the axis
rotates with the field.

\begin{figure}
\centering
\noindent\begin{tabular}{@{}c@{\hspace{0.02\columnwidth}}c@{}}
\subfloat[\label{fig:bead_theta_lab}]{\includegraphics[width=0.47\columnwidth]{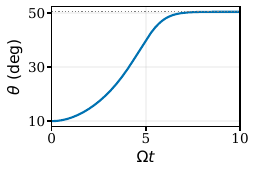}} &
\subfloat[\label{fig:bead_theta_proper}]{\includegraphics[width=0.47\columnwidth]{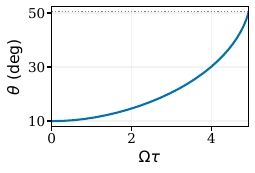}}\\[-1mm]
\subfloat[\label{fig:bead_rho_lab}]{\includegraphics[width=0.47\columnwidth]{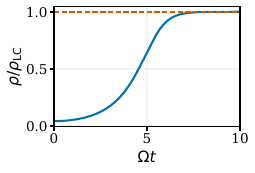}} &
\subfloat[\label{fig:bead_rho_proper}]{\includegraphics[width=0.47\columnwidth]{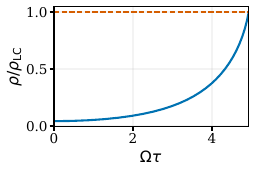}}\\[-1mm]
\subfloat[\label{fig:bead_clock}]{\includegraphics[width=0.47\columnwidth]{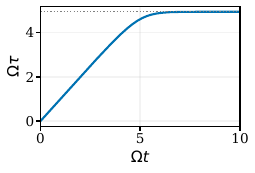}} &
\subfloat[\label{fig:bead_speed}]{\includegraphics[width=0.47\columnwidth]{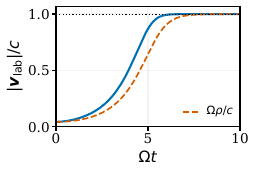}}
\end{tabular}
\caption{Time evolution for the trajectory in Fig.~\ref{fig:bead_path}.
Panels (a)--(d): magnetic angle and distance from the rotation axis against laboratory and proper time. Panel (e): proper time against laboratory time. Panel (f): laboratory speed (solid) and the corotation speed $\Omega\rho$ at the position of the bead (dashed); the dotted line is the speed of light. Horizontal lines mark the angle $\theta_{\rm LC}$ at which the line meets the light cylinder, $\rho=\rho_{\rm LC}$ and the limiting proper time $\tau_{\rm LC}$.}
\label{fig:bead_time}
\end{figure}

Figure~\ref{fig:bead_time} shows the same motion against laboratory
time and against the proper time of the bead. For the first half of the run the bead is slow and the two clocks nearly agree. As the bead approaches the cylinder, $\theta$ and $\rho$ level off in laboratory time, in panels (a) and (c). Plotted against proper time, in panels (b) and (d), they are still rising steeply where the curves end. The proper time itself saturates, as panel (e) shows: it tends to a finite value $\tau_{\rm LC}$, which is about half of the laboratory time covered by the run. The laboratory speed of the bead approaches that of light and stays below it throughout. The dashed curve
$\Omega\rho$ in panel (f) is the speed of a point fixed on the line
at the position of the bead, shown for comparison.

\subsection{The light-cylinder limit}
On this line the first intersection with the light cylinder is at
$\theta_{\rm LC}\simeq50^\circ$. This point is not an equilibrium,
since $C'(\theta_{\rm LC})\ne0$. The only maximum of $\rho$ lies farther along the line, close to the magnetic equator and well outside the cylinder, where a bead fixed on the line would have to move faster than light.

The behaviour near the cylinder follows from the first integral. On the outgoing branch,
\begin{equation}
\dot\theta=\sqrt{\frac{\mathcal E^2-D}{K}},\qquad
\dot t=\frac{\mathcal E+B\dot\theta}{D}.
\end{equation}
At the cylinder $D\to0$ and $K\to B_{\rm LC}^2$. Since $B_{\rm LC}>0$ here,
$\dot\theta\to\mathcal E/B_{\rm LC}$, and hence
\begin{equation}
\frac{d\theta}{dt}\simeq\frac{D}{2B_{\rm LC}},\qquad
\frac{d\tau}{dt}\simeq\frac{D}{2\mathcal E}.
\end{equation}
Close to the cylinder $D\simeq C'_{\rm LC}(\theta_{\rm LC}-\theta)$, so
\begin{equation}
\theta_{\rm LC}-\theta\ \propto\ e^{-\sigma t},
\qquad \sigma=\frac{C'_{\rm LC}}{2B_{\rm LC}}.
\end{equation}
For this line $\sigma\simeq2\Omega$, so the remaining distance to the cylinder shrinks by a factor $e$ in a small fraction of a rotation period. The bead needs an infinite laboratory time to reach the
cylinder, yet only a finite proper time,
$\Omega\tau_{\rm LC}\simeq4.9$ by quadrature. Its laboratory speed tends to $c$, and the induced metric stays regular.

In laboratory time $(d\tau/dt)^2=D-2Bv-Av^2$, with $v=d\theta/dt$, and close to the cylinder the three terms nearly cancel, so a direct integration loses accuracy there. The asymptotic form above does not have this problem.

\subsection{Comparison of motion regimes}
\label{sec:motion_regimes}
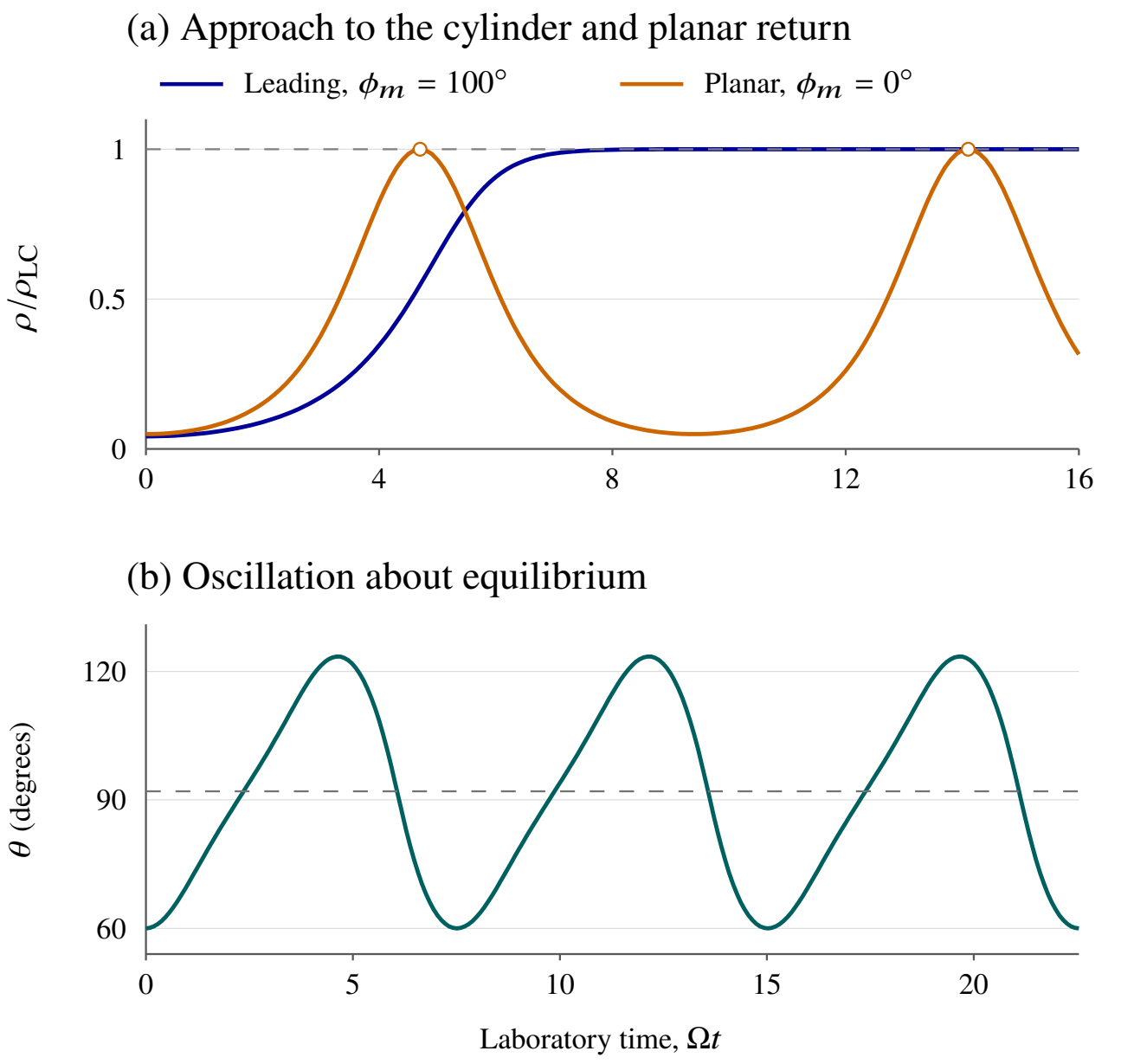
\begin{figure}
\centering
\resizebox{\columnwidth}{!}{\input{bead_regimes_plot}}
\caption{Comparison of three trajectories at $\alpha=45^\circ$ and $R_*=0.01\rho_{\rm LC}$. (a) Cylindrical radius for the approach in Figs.~\ref{fig:bead_path} and \ref{fig:bead_time} and for the planar case with $\phi_m=0$. Open circles mark contacts where the planar solution reaches the speed of light. Both runs start at $\theta_0=10^\circ$ on lines with $R=2\rho_{\rm LC}$. (b) Magnetic polar angle for the bound orbit with $R=0.8\rho_{\rm LC}$ and $\phi_m=100^\circ$; the dashed line marks the stable equilibrium.}
\label{fig:bead_regimes}
\end{figure}

What happens at the light cylinder depends on the sign of $B$ along the direction of travel. Figure~\ref{fig:bead_regimes} shows the leading and planar cases, together with a bound orbit. In every case the bead is released at rest relative to the line.

\textit{Leading line.} This is the trajectory described above. With $B\dot\theta>0$ the line bends forwards along the direction of travel, and the bead winds around the rotation axis as $D\to0$. In Fig.~\ref{fig:bead_regimes}(a) its cylindrical radius tends to $\rho_{\rm LC}$ without reaching it.

\textit{Trailing line.} The laboratory speed obeys $C+2Bv+Av^2<1$. At the cylinder $C=1$, so a massive bead can be there only if $2Bv+Av^2<0$, which requires $Bv<0$: along the direction of travel the line must bend backwards, against the rotation. Replacing $\phi_m=100^\circ$ by $260^\circ$ in the trajectory above leaves $\rho(\theta)$ unchanged and reverses the sign of $B$. On this branch $B\dot\theta\to-\mathcal E$ at the cylinder, which cancels the zero of $D$ in the expression for $\dot t$. The bead crosses the cylinder at a finite laboratory time and at a speed below that of light, in agreement with the role of guide shape found for curved rotating channels \citep{Rogava2003}. In their model of the Crab pulsar outflow, \citet{Gudavadze2015} followed particles beyond the cylinder along three-dimensional field lines that approach an Archimedean spiral.

\textit{Planar line.} Setting $\phi_m=0$, with the other initial data unchanged, gives $B=0$. The first integral reduces to $(d\theta/dt)^2=(D/A)(1-D/\mathcal E^2)$, so the rate along the line vanishes at the cylinder, while the acceleration there, $D'/(2A)$, points back inside. Continued in laboratory time, the solution retraces its path and repeats the outward and inward legs. It is the curved-line counterpart of the oscillatory $\mathrm{dn}$ solution for a straight rotating wire \citep{MachabeliRogava1994,KhomerikiRogava2020}; for the dipole we compute it numerically. At each contact with the cylinder the bead would move at the speed of light and the induced metric degenerates, so the continued solution is not a timelike orbit of a massive bead along its whole length. The planar line is the boundary between leading and trailing lines.

\textit{Bound oscillation.} For $R=0.8\rho_{\rm LC}$, $\phi_m=100^\circ$ and release at $\theta_0=60^\circ$, the whole line lies inside the cylinder and the bead is confined to the well near the magnetic equator. It passes through the equilibrium and turns back on either side where $D=\mathcal E^2$. Because $\mathcal E$ is conserved, the amplitude does not decay. The orbit is timelike throughout.

%% file: bead_regimes_plot.tex
\begin{tikzpicture}[x=1cm,y=1cm,font=\scriptsize]
\draw[black!14,very thin] plot coordinates {(0.900,4.050) (7.550,4.050)};
\node[anchor=east] at (0.860,4.050) {0};
\draw[black!14,very thin] plot coordinates {(0.900,5.118) (7.550,5.118)};
\node[anchor=east] at (0.860,5.118) {0.5};
\draw[black!14,very thin] plot coordinates {(0.900,6.186) (7.550,6.186)};
\node[anchor=east] at (0.860,6.186) {1};
\draw[black!65,thin] plot coordinates {(0.900,4.050) (7.550,4.050)};
\draw[black!65,thin] plot coordinates {(0.900,4.050) (0.900,6.400)};
\node[anchor=north] at (0.900,4.010) {0};
\draw[black!65,thin] plot coordinates {(0.900,4.050) (0.900,4.005)};
\node[anchor=north] at (2.562,4.010) {4};
\draw[black!65,thin] plot coordinates {(2.562,4.050) (2.562,4.005)};
\node[anchor=north] at (4.225,4.010) {8};
\draw[black!65,thin] plot coordinates {(4.225,4.050) (4.225,4.005)};
\node[anchor=north] at (5.888,4.010) {12};
\draw[black!65,thin] plot coordinates {(5.888,4.050) (5.888,4.005)};
\node[anchor=north] at (7.550,4.010) {16};
\draw[black!65,thin] plot coordinates {(7.550,4.050) (7.550,4.005)};
\node[rotate=90] at (0.020,5.210) {$\rho/\rho_{\rm LC}$};
\draw[blue!60!black,line width=.85pt] plot coordinates {(0.900,4.140) (0.942,4.140) (0.983,4.141) (1.025,4.142) (1.066,4.143) (1.108,4.145) (1.149,4.148) (1.191,4.151) (1.232,4.154) (1.274,4.158) (1.316,4.162) (1.357,4.167) (1.399,4.173) (1.440,4.179) (1.482,4.186) (1.523,4.193) (1.565,4.201) (1.607,4.210) (1.648,4.219) (1.690,4.230) (1.731,4.241) (1.773,4.254) (1.814,4.267) (1.856,4.281) (1.898,4.297) (1.939,4.313) (1.981,4.331) (2.022,4.351) (2.064,4.372) (2.105,4.395) (2.147,4.419) (2.188,4.445) (2.230,4.473) (2.272,4.504) (2.313,4.536) (2.355,4.572) (2.396,4.609) (2.438,4.650) (2.479,4.693) (2.521,4.739) (2.562,4.789) (2.604,4.841) (2.646,4.897) (2.687,4.956) (2.729,5.017) (2.770,5.082) (2.812,5.149) (2.853,5.218) (2.895,5.289) (2.937,5.361) (2.978,5.432) (3.020,5.503) (3.061,5.572) (3.103,5.639) (3.144,5.702) (3.186,5.761) (3.228,5.816) (3.269,5.866) (3.311,5.912) (3.352,5.952) (3.394,5.988) (3.435,6.019) (3.477,6.046) (3.518,6.069) (3.560,6.088) (3.602,6.105) (3.643,6.119) (3.685,6.131) (3.726,6.140) (3.768,6.149) (3.809,6.155) (3.851,6.161) (3.893,6.165) (3.934,6.169) (3.976,6.172) (4.017,6.175) (4.059,6.177) (4.100,6.179) (4.142,6.180) (4.183,6.181) (4.225,6.182) (4.267,6.183) (4.308,6.184) (4.350,6.184) (4.391,6.184) (4.433,6.185) (4.474,6.185) (4.516,6.185) (4.558,6.186) (4.599,6.186) (4.641,6.186) (4.682,6.186) (4.724,6.186) (4.765,6.186) (4.807,6.186) (4.848,6.186) (4.890,6.186) (4.932,6.186) (4.973,6.186) (5.015,6.186) (5.056,6.186) (5.098,6.186) (5.139,6.186) (5.181,6.186) (5.223,6.186) (5.264,6.186) (5.306,6.186) (5.347,6.186) (5.389,6.186) (5.430,6.186) (5.472,6.186) (5.513,6.186) (5.555,6.186) (5.597,6.186) (5.638,6.186) (5.680,6.186) (5.721,6.186) (5.763,6.186) (5.804,6.186) (5.846,6.186) (5.888,6.186) (5.929,6.186) (5.971,6.186) (6.012,6.186) (6.054,6.186) (6.095,6.186) (6.137,6.186) (6.178,6.186) (6.220,6.186) (6.262,6.186) (6.303,6.186) (6.345,6.186) (6.386,6.186) (6.428,6.186) (6.469,6.186) (6.511,6.186) (6.553,6.186) (6.594,6.186) (6.636,6.186) (6.677,6.186) (6.719,6.186) (6.760,6.186) (6.802,6.186) (6.843,6.186) (6.885,6.186) (6.927,6.186) (6.968,6.186) (7.010,6.186) (7.051,6.186) (7.093,6.186) (7.134,6.186) (7.176,6.186) (7.218,6.186) (7.259,6.186) (7.301,6.186) (7.342,6.186) (7.384,6.186) (7.425,6.186) (7.467,6.186) (7.508,6.186) (7.550,6.186)};
\draw[black!50,dashed] plot coordinates {(0.900,6.186) (7.550,6.186)};
\draw[orange!80!black,line width=.85pt] plot coordinates {(0.900,4.156) (0.942,4.156) (0.983,4.157) (1.025,4.159) (1.066,4.162) (1.108,4.166) (1.149,4.170) (1.191,4.176) (1.232,4.182) (1.274,4.190) (1.316,4.198) (1.357,4.208) (1.399,4.220) (1.440,4.232) (1.482,4.246) (1.523,4.262) (1.565,4.280) (1.607,4.300) (1.648,4.322) (1.690,4.346) (1.731,4.373) (1.773,4.403) (1.814,4.436) (1.856,4.473) (1.898,4.514) (1.939,4.559) (1.981,4.608) (2.022,4.663) (2.064,4.723) (2.105,4.788) (2.147,4.859) (2.188,4.936) (2.230,5.019) (2.272,5.108) (2.313,5.202) (2.355,5.301) (2.396,5.403) (2.438,5.507) (2.479,5.611) (2.521,5.713) (2.562,5.811) (2.604,5.902) (2.646,5.984) (2.687,6.054) (2.729,6.111) (2.770,6.152) (2.812,6.178) (2.853,6.186) (2.895,6.178) (2.937,6.153) (2.978,6.111) (3.020,6.055) (3.061,5.985) (3.103,5.903) (3.144,5.812) (3.186,5.714) (3.228,5.612) (3.269,5.508) (3.311,5.404) (3.352,5.302) (3.394,5.203) (3.435,5.109) (3.477,5.020) (3.518,4.937) (3.560,4.860) (3.602,4.789) (3.643,4.723) (3.685,4.663) (3.726,4.609) (3.768,4.559) (3.809,4.514) (3.851,4.474) (3.893,4.437) (3.934,4.403) (3.976,4.373) (4.017,4.346) (4.059,4.322) (4.100,4.300) (4.142,4.280) (4.183,4.262) (4.225,4.246) (4.267,4.232) (4.308,4.220) (4.350,4.208) (4.391,4.199) (4.433,4.190) (4.474,4.182) (4.516,4.176) (4.558,4.170) (4.599,4.166) (4.641,4.162) (4.682,4.159) (4.724,4.157) (4.765,4.156) (4.807,4.156) (4.848,4.156) (4.890,4.157) (4.932,4.159) (4.973,4.162) (5.015,4.166) (5.056,4.170) (5.098,4.176) (5.139,4.182) (5.181,4.190) (5.223,4.198) (5.264,4.208) (5.306,4.219) (5.347,4.232) (5.389,4.246) (5.430,4.262) (5.472,4.280) (5.513,4.299) (5.555,4.321) (5.597,4.346) (5.638,4.373) (5.680,4.403) (5.721,4.436) (5.763,4.473) (5.804,4.514) (5.846,4.559) (5.888,4.608) (5.929,4.662) (5.971,4.722) (6.012,4.787) (6.054,4.859) (6.095,4.936) (6.137,5.019) (6.178,5.107) (6.220,5.201) (6.262,5.300) (6.303,5.402) (6.345,5.506) (6.386,5.610) (6.428,5.713) (6.469,5.811) (6.511,5.902) (6.553,5.983) (6.594,6.054) (6.636,6.110) (6.677,6.152) (6.719,6.178) (6.760,6.186) (6.802,6.178) (6.843,6.153) (6.885,6.112) (6.927,6.055) (6.968,5.985) (7.010,5.904) (7.051,5.813) (7.093,5.715) (7.134,5.613) (7.176,5.509) (7.218,5.404) (7.259,5.302) (7.301,5.204) (7.342,5.110) (7.384,5.021) (7.425,4.937) (7.467,4.860) (7.508,4.789) (7.550,4.724)};
\draw[orange!80!black,fill=white] (2.854,6.186) circle (1.3pt);
\draw[orange!80!black,fill=white] (6.761,6.186) circle (1.3pt);
\node[anchor=west,font=\small] at (0.660,7.030) {(a) Approach to the cylinder and planar return};
\draw[blue!60!black,thick] plot coordinates {(1.000,6.650) (1.450,6.650)};
\node[anchor=west] at (1.500,6.650) {Leading, $\phi_m=100^\circ$};
\draw[orange!80!black,thick] plot coordinates {(4.280,6.650) (4.730,6.650)};
\node[anchor=west] at (4.780,6.650) {Planar, $\phi_m=0^\circ$};
\draw[black!14,very thin] plot coordinates {(0.900,0.633) (7.550,0.633)};
\node[anchor=east] at (0.860,0.633) {60};
\draw[black!14,very thin] plot coordinates {(0.900,1.549) (7.550,1.549)};
\node[anchor=east] at (0.860,1.549) {90};
\draw[black!14,very thin] plot coordinates {(0.900,2.464) (7.550,2.464)};
\node[anchor=east] at (0.860,2.464) {120};
\draw[black!65,thin] plot coordinates {(0.900,0.450) (7.550,0.450)};
\draw[black!65,thin] plot coordinates {(0.900,0.450) (0.900,2.800)};
\node[anchor=north] at (0.900,0.410) {0};
\draw[black!65,thin] plot coordinates {(0.900,0.450) (0.900,0.405)};
\node[anchor=north] at (2.375,0.410) {5};
\draw[black!65,thin] plot coordinates {(2.375,0.450) (2.375,0.405)};
\node[anchor=north] at (3.851,0.410) {10};
\draw[black!65,thin] plot coordinates {(3.851,0.450) (3.851,0.405)};
\node[anchor=north] at (5.326,0.410) {15};
\draw[black!65,thin] plot coordinates {(5.326,0.450) (5.326,0.405)};
\node[anchor=north] at (6.801,0.410) {20};
\draw[black!65,thin] plot coordinates {(6.801,0.450) (6.801,0.405)};
\node[rotate=90] at (0.020,1.610) {$\theta$ (degrees)};
\draw[teal!75!black,line width=.85pt] plot coordinates {(0.900,0.633) (0.928,0.637) (0.955,0.647) (0.983,0.663) (1.011,0.685) (1.039,0.713) (1.066,0.745) (1.094,0.782) (1.122,0.822) (1.149,0.865) (1.177,0.910) (1.205,0.958) (1.232,1.006) (1.260,1.054) (1.288,1.103) (1.316,1.152) (1.343,1.200) (1.371,1.247) (1.399,1.294) (1.426,1.340) (1.454,1.385) (1.482,1.430) (1.510,1.474) (1.537,1.518) (1.565,1.562) (1.593,1.605) (1.620,1.649) (1.648,1.692) (1.676,1.737) (1.704,1.781) (1.731,1.826) (1.759,1.872) (1.787,1.919) (1.814,1.967) (1.842,2.015) (1.870,2.064) (1.898,2.114) (1.925,2.164) (1.953,2.213) (1.981,2.263) (2.008,2.310) (2.036,2.356) (2.064,2.400) (2.091,2.439) (2.119,2.475) (2.147,2.506) (2.175,2.532) (2.202,2.551) (2.230,2.564) (2.258,2.570) (2.285,2.570) (2.313,2.561) (2.341,2.545) (2.369,2.521) (2.396,2.489) (2.424,2.448) (2.452,2.399) (2.479,2.341) (2.507,2.274) (2.535,2.198) (2.563,2.112) (2.590,2.017) (2.618,1.913) (2.646,1.801) (2.673,1.683) (2.701,1.563) (2.729,1.443) (2.756,1.327) (2.784,1.219) (2.812,1.120) (2.840,1.030) (2.867,0.951) (2.895,0.881) (2.923,0.820) (2.950,0.769) (2.978,0.726) (3.006,0.692) (3.034,0.666) (3.061,0.647) (3.089,0.637) (3.117,0.633) (3.144,0.637) (3.172,0.647) (3.200,0.663) (3.228,0.685) (3.255,0.713) (3.283,0.745) (3.311,0.782) (3.338,0.822) (3.366,0.865) (3.394,0.910) (3.421,0.958) (3.449,1.006) (3.477,1.054) (3.505,1.103) (3.532,1.152) (3.560,1.200) (3.588,1.247) (3.615,1.294) (3.643,1.340) (3.671,1.385) (3.699,1.430) (3.726,1.474) (3.754,1.518) (3.782,1.562) (3.809,1.605) (3.837,1.649) (3.865,1.692) (3.893,1.737) (3.920,1.781) (3.948,1.826) (3.976,1.872) (4.003,1.919) (4.031,1.967) (4.059,2.015) (4.086,2.064) (4.114,2.114) (4.142,2.164) (4.170,2.213) (4.197,2.263) (4.225,2.310) (4.253,2.356) (4.280,2.400) (4.308,2.439) (4.336,2.475) (4.364,2.506) (4.391,2.532) (4.419,2.551) (4.447,2.564) (4.474,2.570) (4.502,2.570) (4.530,2.561) (4.558,2.545) (4.585,2.521) (4.613,2.489) (4.641,2.448) (4.668,2.399) (4.696,2.341) (4.724,2.274) (4.751,2.198) (4.779,2.112) (4.807,2.017) (4.835,1.913) (4.862,1.801) (4.890,1.683) (4.918,1.563) (4.945,1.443) (4.973,1.327) (5.001,1.219) (5.029,1.120) (5.056,1.030) (5.084,0.951) (5.112,0.881) (5.139,0.820) (5.167,0.769) (5.195,0.726) (5.223,0.692) (5.250,0.666) (5.278,0.647) (5.306,0.637) (5.333,0.633) (5.361,0.637) (5.389,0.647) (5.416,0.663) (5.444,0.685) (5.472,0.713) (5.500,0.745) (5.527,0.782) (5.555,0.822) (5.583,0.865) (5.610,0.910) (5.638,0.958) (5.666,1.006) (5.694,1.054) (5.721,1.103) (5.749,1.152) (5.777,1.200) (5.804,1.247) (5.832,1.294) (5.860,1.340) (5.888,1.385) (5.915,1.430) (5.943,1.474) (5.971,1.518) (5.998,1.562) (6.026,1.605) (6.054,1.649) (6.081,1.692) (6.109,1.737) (6.137,1.781) (6.165,1.826) (6.192,1.872) (6.220,1.919) (6.248,1.967) (6.275,2.015) (6.303,2.064) (6.331,2.114) (6.359,2.164) (6.386,2.213) (6.414,2.263) (6.442,2.310) (6.469,2.356) (6.497,2.400) (6.525,2.439) (6.553,2.475) (6.580,2.506) (6.608,2.532) (6.636,2.551) (6.663,2.564) (6.691,2.570) (6.719,2.570) (6.746,2.561) (6.774,2.545) (6.802,2.521) (6.830,2.489) (6.857,2.448) (6.885,2.399) (6.913,2.341) (6.940,2.274) (6.968,2.198) (6.996,2.112) (7.024,2.017) (7.051,1.913) (7.079,1.801) (7.107,1.683) (7.134,1.563) (7.162,1.443) (7.190,1.327) (7.218,1.219) (7.245,1.120) (7.273,1.030) (7.301,0.951) (7.328,0.881) (7.356,0.820) (7.384,0.769) (7.411,0.726) (7.439,0.692) (7.467,0.666) (7.495,0.647) (7.522,0.637) (7.550,0.633)};
\draw[black!55,dashed] plot coordinates {(0.900,1.610) (7.550,1.610)};
\node[anchor=west,font=\small] at (0.660,3.120) {(b) Oscillation about equilibrium};
\node[] at (4.150,-0.170) {Laboratory time, $\Omega t$};
\end{tikzpicture}

%% file: 06_numerical_results.tex
\section{Numerical equilibrium maps}
\label{sec:numerical_equilibria}

\subsection{Construction of the maps}
\begin{figure*}
\centering
\includegraphics[width=\textwidth]{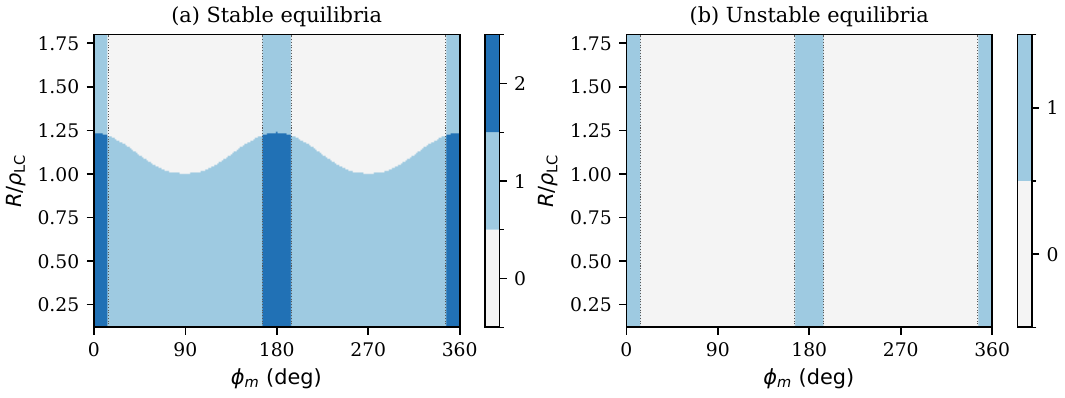}
\caption{Number of allowed stable (left) and unstable (right)
equilibria per dipolar line for $\alpha=45^\circ$ and
$R_*=0.01\rho_{\rm LC}$. Only positions outside the star and inside the light
cylinder are counted. Dotted vertical lines mark the mergers of
stationary points. The curved boundary in the left panel marks
where the more distant stable branch leaves the light cylinder.
The stellar surface removes no stationary point over this
parameter range.}
\label{fig:equilibrium_counts}
\end{figure*}

We now map the equilibria for one intermediate inclination, $\alpha=45^\circ$, with $R_*=0.01\rho_{\rm LC}$. The equatorial radius is scanned over $0.12\leq R/\rho_{\rm LC}\leq1.8$, which includes lines that close inside the light cylinder and lines that cross it, and the magnetic azimuth over the full circle. At each azimuth we solve the cubic, add
the equatorial root where the tangent substitution misses it, and
classify each root by the sign of $C''$. A position is kept if it lies outside the star and inside the light cylinder. Figure~\ref{fig:equilibrium_counts} shows the resulting numbers of stable and unstable equilibria on each line.

\subsection{Where the branches end}
The maps are easiest to read starting from the tilt plane. At $\phi_m=0$ the line has the three stationary points of Section~\ref{sec:special_cases}. The middle one is the axis crossing at $\theta=\pi-\alpha$ and is unstable. The other two are stable. The northern one is not far from the magnetic equator, and its distance from the rotation axis is a large fraction of $R$. The southern one lies between the axis crossing and the southern footpoint and stays close to the rotation axis, at less than a tenth of $R$.

As $\phi_m$ increases, the line no longer crosses the axis. The unstable point becomes a minimum of $\rho$ at a small distance from the axis and moves towards the nearby stable point. The two merge where $C'=C''=0$, which for this inclination happens at an azimuth $\phi_{\rm c}\simeq14^\circ$. The merging pair is close to the rotation axis, so the merger takes place well inside the light cylinder for every $R$ in the scanned range. Reflection gives the other three merger azimuths. Within $\phi_{\rm c}$ of either $0^\circ$ or $180^\circ$ a line has two maxima and one minimum of $\rho$; elsewhere it has a single maximum.

\begin{figure}
\centering
\includegraphics[width=\columnwidth]{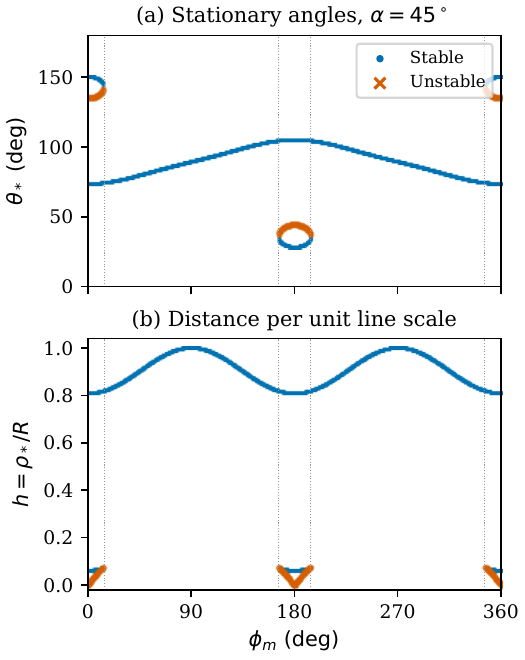}
\caption{Stationary branches at $\alpha=45^\circ$:
magnetic polar angle (top) and $h=\rho_*/R$ (bottom). Blue points
are stable and orange crosses unstable. The branches are shown
before the stellar-surface and light-cylinder cuts; their angles
and $h$ values are independent of $R$. Dotted lines mark the four
merger azimuths. In the lower panel the nearby stable and unstable
branches have very similar distance close to a merger.}
\label{fig:equilibrium_branches}
\end{figure}

Figure~\ref{fig:equilibrium_branches} shows the stationary
angles and their values of $h=\rho_*/R$. The stable branch that exists at
all azimuths stays within about $15^\circ$ of the magnetic equator, with $h$ between about $0.8$ and unity. It leaves the light
cylinder when $R$ exceeds
$R_{\rm LC}(\phi_m)=1/(\Omega h)$. This limit equals $\rho_{\rm LC}$ for lines perpendicular to the tilt plane, where the maximum of $\rho$ is at the magnetic equator, and it is about a quarter larger for lines in the tilt plane. Above it the maximum of $\rho$ still exists but lies outside the light cylinder.
The stable--unstable pair near the axis stays inside the light cylinder
over the whole scanned range, and the stellar surface removes no
stationary point for these parameters.
In the count maps of Fig.~\ref{fig:equilibrium_counts}, the vertical edges are therefore mergers and the curved edge is the crossing of the light cylinder.

\subsection{Spatial distribution}
\begin{figure}
\centering
\includegraphics[width=\columnwidth]{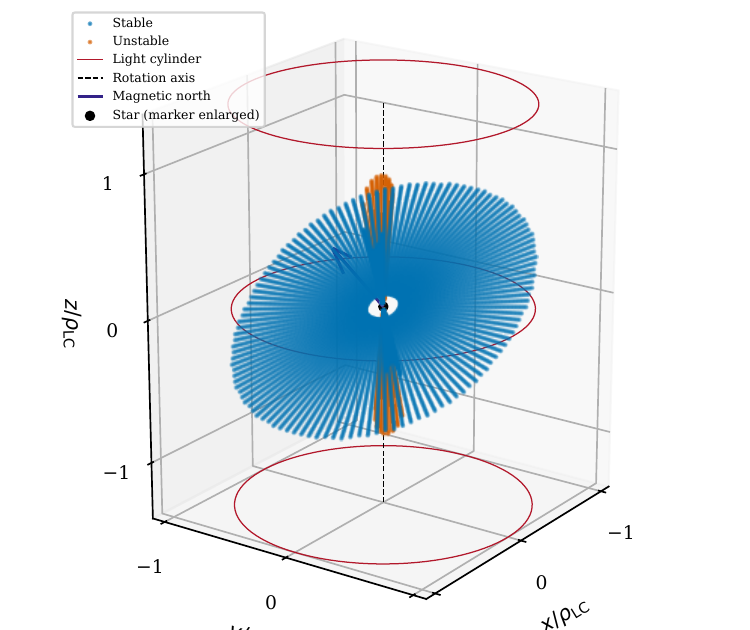}
\caption{Spatial distribution of the allowed equilibria at $t=0$. Blue and orange points denote stable and
unstable positions, respectively. Red circles outline the light
cylinder, the dashed line is the rotation axis, and the arrow marks
the magnetic north direction. The star is drawn enlarged; its radius is $0.01\rho_{\rm LC}$. Lengths are in units of $\rho_{\rm LC}$.}
\label{fig:equilibrium_space}
\end{figure}

Figure~\ref{fig:equilibrium_space} shows the allowed positions in
laboratory space at $t=0$. For a given stationary angle, changing
$R$ moves the point along a straight line through the origin. The points trace out portions of equilibrium surfaces, which
rotate rigidly with the star: at a later time the whole pattern is
turned about the rotation axis, with the counts, polar angles and
distances from the axis unchanged.

%% file: 08_tilt_dependence.tex
\section{How the number of wells depends on inclination}
\label{sec:tilt_concentration}

\subsection{A normalized measure}
To compare inclinations we use the same set of field lines at every
$\alpha$. Let $N_{\rm s}(\alpha,R,\phi_m)$ be the number of stable
equilibria on a line that lie outside the star and inside the light
cylinder. Its mean over the set is
\begin{equation}
\overline N_{\rm s}(\alpha)=
\frac{1}{2\pi}\int_0^{2\pi}\!d\phi_m
\int_{R_{\min}}^{R_{\max}}\!w(R)
N_{\rm s}(\alpha,R,\phi_m)\,dR,
\label{eq:tilt_mean_stable}
\end{equation}
where $\int_{R_{\min}}^{R_{\max}}w(R)\,dR=1$.
As a reference we weight the lines uniformly in equatorial radius,
$w(R)=1/(R_{\max}-R_{\min})$, with
\begin{equation}
R_*=0.01\rho_{\rm LC},\qquad
0.12\leq R/\rho_{\rm LC}\leq1.8.
\label{eq:tilt_ensemble}
\end{equation}
This interval contains both lines that close inside the cylinder and lines that intersect it. It serves to compare tilts and makes no claim about the distribution of footpoints on a real pulsar. The azimuth is uniform on $[0,2\pi)$ and $\Omega$ is fixed.

Because a line can hold two stable points, $\overline N_{\rm s}$
can exceed unity. We also give $P_{\geq1}$, the
fraction of lines with at least one stable point, obtained by
replacing $N_{\rm s}$ in Eq.~\eqref{eq:tilt_mean_stable} with the
indicator $\boldsymbol{1}_{N_{\rm s}\geq1}$. If $P_2$ is the fraction
with two stable points, then $\overline N_{\rm s}=P_{\geq1}+P_2$.

\subsection{Calculation and tilt dependence}
For each $(\alpha,\phi_m)$ we find the stable stationary angles as in
Section~\ref{sec:general_equilibria}. The integral over $R$ can then
be done analytically. A stable angle $\theta_i$, with
$h_i=\rho_i/R$, is allowed for $L_i<R<U_i$, where
\begin{align}
L_i&=\max\left(R_{\min},\frac{R_*}{\sin^2\theta_i}\right),\\
U_i&=\min\left(R_{\max},\frac{1}{\Omega h_i}\right).
\end{align}
For uniform weighting it contributes
$[U_i-L_i]_+/(R_{\max}-R_{\min})$ to the azimuthal average, where
$[x]_+=\max(x,0)$. If $h_i=0$, the upper bound from the light
cylinder is absent.

\begin{figure}
\centering
\includegraphics[width=\columnwidth]{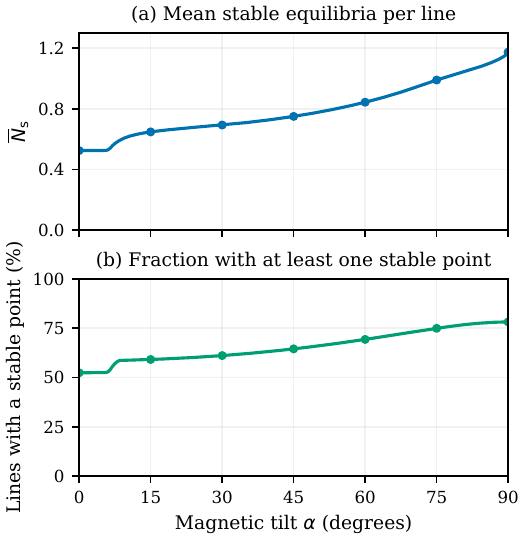}
\caption{Tilt dependence for the field-line ensemble of Eq.~\eqref{eq:tilt_ensemble}, with uniform weighting in $R$ and magnetic azimuth. Top: mean number of stable equilibria per line that lie outside the star and inside the light cylinder. Bottom: percentage of lines with at least one such point.}
\label{fig:tilt_simple}
\end{figure}

Figure~\ref{fig:tilt_simple} shows the result. The mean count increases
monotonically with tilt, from about one half for the aligned rotator to somewhat more than one for the orthogonal rotator. Apart from a step at small inclination, which is discussed below, the growth is gradual and becomes faster towards orthogonality. Two things contribute to it. More lines have a stable point at larger tilt: $P_{\geq1}$ rises from about one half to more than three quarters. Some lines also have two. Such lines are absent at alignment and make up more than a third of the sample at $\alpha=90^\circ$.

The size of the increase depends on the interval of $R$. Raising $R_{\max}$ with $R_{\min}$ held fixed adds long lines, which have no stable point inside the cylinder in the aligned case but can have one when the magnetic axis is tilted. The ratio of the orthogonal to the aligned count therefore grows with $R_{\max}$. It is well below two for an interval that ends just outside the light cylinder and between two and three for longer ones. The ratio stops changing once $R_{\max}$ exceeds $(3\sqrt{3}/2)\rho_{\rm LC}$, because longer lines have no stable point at either end of the inclination range and only alter the common normalization. The count always increases with tilt, but by an amount that depends on $R_{\max}$.

\subsection{Interpretation and dependence on the weighting}
The aligned value can be understood directly. The only stable point
is at the magnetic equator, where $\rho=R$, so only lines with
$R<\rho_{\rm LC}$ contribute and
\begin{equation}
\overline N_{\rm s}(0)
=\frac{\rho_{\rm LC}-R_{\min}}{R_{\max}-R_{\min}},
\end{equation}
which is close to one half for the reference interval.
As the tilt increases, the range of azimuths
with two stable points widens, and some lines with
$R>\rho_{\rm LC}$ acquire stable points whose distance from the
rotation axis is still less than $\rho_{\rm LC}$. At small
inclination the additional branch lies inside the star, and it emerges over a narrow range of tilt. This
produces the step at low inclination in
Fig.~\ref{fig:tilt_simple}, whose position shifts with the stellar
radius.

Uniform weighting in $R$ is convenient but not unique. Weighting by
magnetic flux through the northern stellar footpoints gives
$d\Phi\propto R^{-2}dR\,d\phi_m$ for a dipole, and the normalized
weight
\begin{equation}
w_\Phi(R)=\frac{R^{-2}}{R_{\min}^{-1}-R_{\max}^{-1}}.
\end{equation}
With this weight the mean stable count still rises with tilt, but only by about one half between the aligned and orthogonal cases, where uniform weighting gives more than a factor of two. The increase is smaller
because flux weighting favours short lines, which already
have a stable point in the aligned case.

\subsection{Where a stable point appears and disappears}
\label{sec:stable_radius_window}

The bounds in Eq.~\eqref{eq:arbitrary_phi_R_interval} show how a stable point comes and goes as the line is made longer: it emerges from the star at the lower bound and leaves through the light cylinder at the upper bound. A useful example is the southern branch in the tilt plane ($\phi_m=0$) at small positive $\alpha$. There $\pi-\theta_i\simeq2\alpha/3$ and $h_i\simeq4\alpha^3/27$, so the lower and upper bounds scale as $9R_*/(4\alpha^2)$ and $27\rho_{\rm LC}/(4\alpha^3)$, respectively, with $\alpha$ in radians. At any fixed nonzero tilt, a sufficiently long line has no stable point inside the cylinder. At fixed $R>\rho_{\rm LC}$, a sufficiently small tilt also removes them: the southern point is then inside the star, and the branch that tends to the magnetic equator is outside the cylinder. Shorter lines keep their stable points near alignment.

\begin{figure}
\centering
\makebox[\columnwidth][l]{\small (a) First stable point as $R$ decreases}\par
\includegraphics[width=\columnwidth]{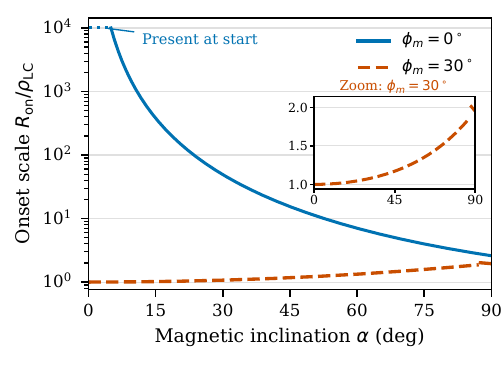}\par\smallskip
\makebox[\columnwidth][l]{\small (b) Fixed equatorial radius}\par
\includegraphics[width=\columnwidth]{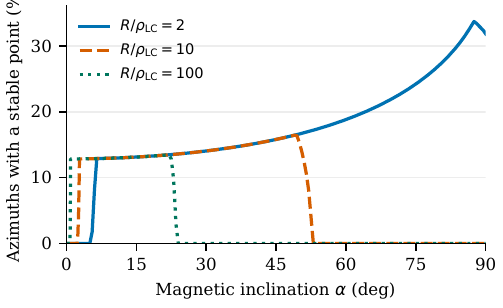}
\caption{Stable equilibria outside the star and inside the light cylinder, for $R_*=0.01\rho_{\rm LC}$. (a) The equatorial radius just below which the first stable point is allowed when $R$ is decreased from $10^4\rho_{\rm LC}$. The dotted segment near alignment marks cases where a stable point is already present at the starting radius, and the inset shows the $\phi_m=30^\circ$ curve on a linear scale. (b) Percentage of magnetic azimuths with at least one stable point at three fixed equatorial radii.}
\label{fig:fixed_radius_equilibria}
\end{figure}

Figure~\ref{fig:fixed_radius_equilibria}(a) shows the result of decreasing $R$ from a very large value at each inclination and azimuth and recording the first equatorial radius at which a stable point lies between the star and the cylinder. In the tilt plane the onset is set by the stable point near the rotation axis and follows the steep scaling above. Close to alignment such a point exists on lines thousands of times longer than $\rho_{\rm LC}$, and towards orthogonality the onset falls to a few $\rho_{\rm LC}$. The line in the tilt plane can thus have a stable point close to the rotation axis even when $R$ is large. Away from the tilt plane the onset is much lower. At $\phi_m=30^\circ$ there is a single stable branch at almost all inclinations, and the onset stays between one and about two $\rho_{\rm LC}$. The small jump in this curve close to orthogonality is where a second stable point and an unstable one appear. Below the onset a stable point need not persist at all smaller $R$, because each branch has its own allowed interval.

Panel (b) gives the fraction of azimuths with a stable point at three fixed equatorial radii, all larger than $\rho_{\rm LC}$. The three curves have the same structure. Close to alignment the fraction is zero, because the stable points near the rotation axis are still inside the star. Once they emerge, the fraction follows a curve common to the three radii. It rises slowly with inclination, as the range of azimuths about the tilt plane with a second maximum of $\rho$ widens. The fraction then drops to zero at the inclination where these points move outside the light cylinder. The longer the line, the smaller this inclination: on the longest line shown stable points exist only within about $25^\circ$ of alignment, and on the shortest they persist almost up to orthogonality.

\subsection{Wells on the route to the first cylinder crossing}
\label{sec:footpoint_supply}

A well that lies beyond the light cylinder along the line cannot stop a particle on its way out from a given pole. For the supply to the cylinder, what matters is whether a well lies on the way. We count directed paths. For each $(R,\phi_m)$ we take the two stellar footpoints with equal weight and follow the line from each towards the other until $\rho=\rho_{\rm LC}$ for the first time. Let $I_{\rm LC}$ mark a path that reaches the cylinder, and $I_{\rm pre}$ such a path with a stable point strictly between the launch and the first contact. We define the conditional fraction
\begin{equation}
 F_{\rm pre}(\alpha)=\frac{\langle I_{\rm pre}\rangle}
                              {\langle I_{\rm LC}\rangle}.
\label{eq:footpoint_fraction}
\end{equation}
The averages use the same weights in $R$ and azimuth as before, and a line that never reaches the cylinder contributes to neither. The fraction describes the geometry of the path only; whether a massive bead actually crosses the cylinder is the separate question treated in Section~\ref{sec:motion_regimes}.

The numerator is simple to evaluate. Order the local maxima of $\rho$ from the chosen footpoint. The first one with $Rh_i\geq\rho_{\rm LC}$ guarantees a crossing on its near side, since the stellar surface is entirely inside the cylinder. The path counts if an earlier maximum lies outside the star and inside the cylinder. A line has at most two maxima, so this requires a lower peak on the launch side and a higher peak, reaching the cylinder, farther along. Only one of the two footpoints of such a line then contributes to the numerator. With Eq.~\eqref{eq:arbitrary_phi_R_interval}, the integral over $R$ can again be done before the average over azimuth.

At $45^\circ$, somewhat fewer than half of the reference lines reach the cylinder. Only about one in fifteen of their directed paths meets a well first, although roughly two thirds of \emph{all} lines have a well somewhere. The conditional fraction vanishes at both ends of the inclination range: at alignment a line has a single maximum of $\rho$, and at orthogonality the two maxima, when present, have equal height.

Much longer lines, $100<R/\rho_{\rm LC}<1000$, all reach the cylinder, and their wells are confined to small inclinations. The fraction of these lines with a well has a broad, nearly level maximum around $10^\circ$, where it is a little over one tenth; it is too flat there to single out an angle. Only one of the two footpoints of such a line has the well on its route, so about half that fraction of the directed paths meets a well first. Beyond an inclination of about $25^\circ$ none does.

Whether a bead gets past a well depends on its energy. Let $D_{\rm b}=\max D$ on the segment between the launch point and a target just inside the cylinder. An outgoing bead with $\mathcal E^2>D_{\rm b}$ passes every barrier on the way; equality at an interior maximum puts it on a separatrix. A bead that has entered a well across a barrier is not captured, because without dissipation it can leave the same way. A well can, however, hold particles that are created inside it with less than the escape energy.

%% file: 07_equilibrium_geometry.tex
\section{Limits of the prescribed field}
\label{sec:discussion}

The results so far are statements about geometry. Before using them to say anything about pulsars, we set out what the model leaves out.

 The bead follows an imposed, rigidly rotating dipole. This isolates the centrifugal force along a curved line, but it supplies neither a parallel electric field nor the plasma currents that set the real shape of the line. Even in vacuum the electromagnetic field of an inclined dipole is retarded \citep{Deutsch1955}. In plasma-filled solutions the open field bends back near the light cylinder and joins the wind \citep{Contopoulos1999,Spitkovsky2006}. Our label $R$ is the equatorial radius of a static dipolar shape. It does not tell which lines of a real oblique magnetosphere are open and which are closed.

 The static dipole is least reliable for the wells on long lines. The light cylinder is a limit on the cylindrical radius $\rho$, so a point can be inside it and still far from the star in spherical distance $r$. On the near-axis branch in the tilt plane at $\alpha=10^\circ$, for example, the well sits at a spherical radius $r=R\sin^2\theta_*$ of about one per cent of $R$. Over the long-line sample this places it between roughly one and ten $\rho_{\rm LC}$ from the star, even though $\rho<\rho_{\rm LC}$. A static dipole extrapolated this far up the rotation axis cannot establish a preferred inclination for real pulsars.

 Close to the star, gravity matters. For a star of mass $M$, the Newtonian gravitational and centrifugal accelerations are equal in the equatorial plane at $r_{\rm co}=(GM/\Omega^2)^{1/3}$, or $r_{\rm co}/\rho_{\rm LC}=(GM\Omega/c^3)^{1/3}$. For $1.4\,M_\odot$ this is about a tenth at the Crab period and a few per cent at $P=1$ s. Inside this radius gravity exceeds the centrifugal acceleration in the equatorial plane; elsewhere the components of the two forces along the particular line must be compared.

 The strongest restriction comes from energy. A bead launched inside the cylinder has $\mathcal E=\sqrt{D_0}\,\gamma_{\parallel,0}$, where $\gamma_{\parallel,0}$ is the Lorentz factor measured by the local corotating observer. Bounded oscillation requires $\mathcal E<1$, which means $|V_{\parallel,0}|<\Omega\rho_0$ in our units: the bead must be launched along the line more slowly than the line itself moves at that point. At a polar-cap footpoint, where $\rho_0\ll\rho_{\rm LC}$, this is a severe restriction. Polar-cap cascades produce secondary pairs, which can already have Lorentz factors of order $10^2$, and primaries accelerated to factors of order $10^6$--$10^7$ \citep{DaughertyHarding1982,TimokhinHarding2015}. A relativistic outflow along the field from $\rho_0\ll\rho_{\rm LC}$ therefore passes every well of the model. The wells can be populated in other ways. Close to the cylinder $D_0$ is small, so a large Lorentz factor in the laboratory does not by itself put a particle above the threshold. Particles can be created slowly inside a well. Dissipation can also change the energy and even the equilibria themselves, as in the photon-drag calculation of \citet{Bakhtadze2019}.

 A charge that merely corotates in a well radiates little. At an exact equilibrium its laboratory orbit is a circle. Restoring $c$ and using Gaussian units, the power radiated by a single electron through its transverse acceleration is
\begin{equation}
 P_{\rm cor}=\frac{2e^2\Omega^2}{3c}\,\beta_{\rm cor}^2\gamma_{\rm cor}^4,
 \qquad \beta_{\rm cor}=\rho/\rho_{\rm LC}.
\end{equation}
At the Crab period the prefactor is of order $10^{-25}\ {\rm erg\,s^{-1}}$, so for modest $\gamma_{\rm cor}$ the power per electron is very small. The formal Lorentz factor rises close to the cylinder, but that is where the assumptions of a rigid line and no losses are least secure. A count of wells supplies neither a particle density nor an energy source for a bright trapped component.

With these limits in mind, the next section asks only whether measured inclinations can distinguish the geometrical trend.

%% file: 08b_observed_inclinations.tex
\section{Can measured inclinations test the trapping geometry?}
\label{sec:observed_inclinations}

\begin{figure*}
\centering
\includegraphics[width=\textwidth]{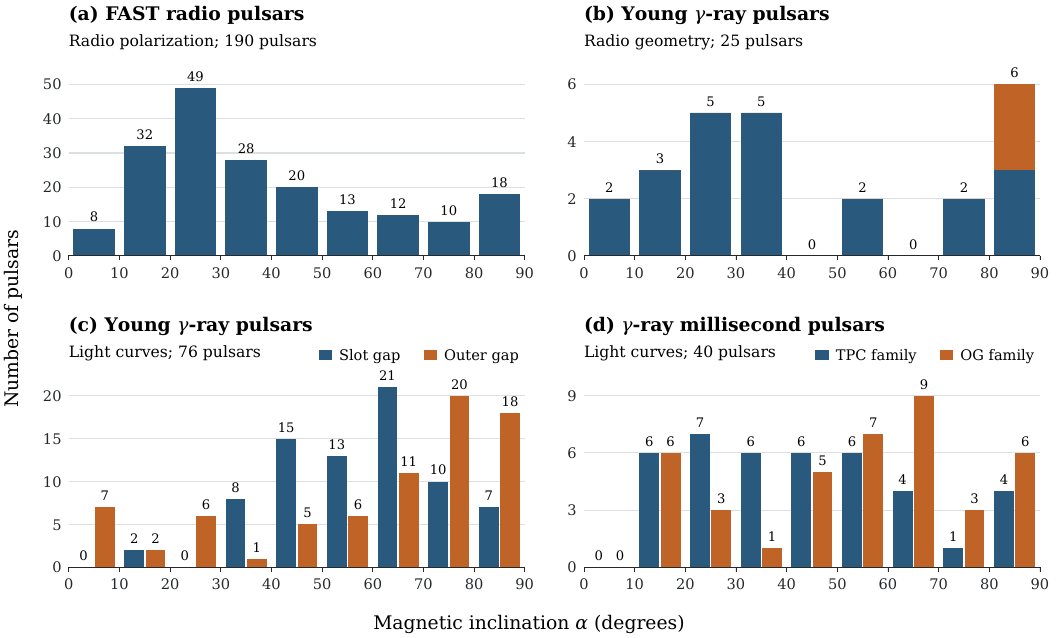}
\caption{Number of pulsars per $10^\circ$ bin of folded magnetic inclination. Each bar is labelled with its count; bins include their lower edge and the final bin includes $90^\circ$. (a) FAST radio polarization fits \citep[table~5]{WangFAST2023}. (b) Preferred radio geometries for 25 of 28 young $\gamma$-ray pulsars \citep[table~2]{Rookyard2015}; the orange part of the last bar marks three assumed orthogonal geometries. (c) Slot-gap and outer-gap light-curve fits for the same 76 young pulsars \citep[tables~1 and~3]{PierbattistaGeom2015}. (d) Two-pole caustic (TPC) and outer-gap (OG) model families for the same 40 millisecond pulsars \citep[tables~6--9 and~11]{JohnsonMSP2014}. Both include their altitude-limited variants and share six pair-starved polar-cap fits. Paired bars are alternative fits, and samples can overlap between panels.}
\label{fig:obs_angle_counts}
\end{figure*}

\subsection{A simple score for outer emission}
\label{sec:emission_score}

To compare the model with pulsar samples we need a quantity that depends on inclination and is at least loosely related to what is observed. We construct the simplest one, for emission from near the light cylinder.

Two effects compete in the idealized picture. A line that supports a well might supply fewer particles to an accelerator near the light cylinder. Increasing the tilt, on the other hand, brings the cylinder crossing of a nearly axial line closer to the star, where the dipole field is stronger. We give each sampled line the same supply, assign zero outer yield to lines with wells, and weight the remaining lines by $\sin^p\alpha$. The resulting dimensionless score is
\begin{equation}
 S_p(\alpha)=[1-P_{\geq1}(\alpha)]\sin^p\!\alpha .
 \label{eq:inclination_emission_proxy}
\end{equation}
The linear weight ($p=1$) is a simple reference choice. The cubic weight ($p=3$) corresponds to a yield proportional to the field strength on a nearly axial line: the crossing is at $r\simeq\rho_{\rm LC}/\sin\alpha$, and $|\boldsymbol B|\propto r^{-3}$. This estimate applies when the angular displacement of the line from the magnetic axis at the crossing is small compared with $\alpha$.

For the ensemble in Eq.~\eqref{eq:tilt_ensemble}, the score is largest near $60^\circ$ for $p=1$ and near $75^\circ$ for $p=3$. Both maxima are broad: in each case the score stays within a few per cent of its maximum over an interval about $25^\circ$ wide. For a different range of lines, $1<R/\rho_{\rm LC}<10$, both maxima move to orthogonality.

The score says little about trapping. In this ensemble every line without a well reaches the cylinder, so $1-P_{\geq1}$ is close to the fraction of lines that reach it. The fall of this factor towards orthogonality mostly reflects that fewer lines of a fixed range of $R$ reach the cylinder as the tilt grows. Counting only wells on the route to the cylinder, as in Eq.~\eqref{eq:footpoint_fraction}, shifts the maxima by a few degrees. The score also treats every well as a perfect trap, which Section~\ref{sec:discussion} shows it is not for energetic particles, and a luminosity would require the routes to be weighted by the particles that actually travel along them. What can usefully be compared with observations is a range of favoured inclinations, roughly $45^\circ$ to $90^\circ$; the score does not predict a peak.

\subsection{What the radio samples measure}

A count of field lines refers to one star of given inclination. A pulsar catalogue counts stars, and the inclination of each must first be inferred. We identify opposite magnetic poles and quote angles in the range $0\leq\alpha\leq90^\circ$. For randomly oriented magnetic axes the probability density in this convention is $\sin\alpha$ per radian, because equal intervals of inclination do not subtend equal solid angles.

Radio polarimetry shows how hard such a catalogue is to build. Near the inflection point of the polarization position angle $\psi$, the rotating-vector model gives $|{\rm d}\psi/{\rm d}\varphi|=|\sin\alpha/\sin\beta|$, where $\varphi$ is pulse longitude and $\beta$ is the closest angular approach of the sightline to the magnetic pole. A short stretch of the swing therefore constrains only a ratio of the two angles. Of the young pulsars studied by \citet{JohnstonWeisberg2006}, only PSR~J1015$-$5719 constrained both angles tightly. Its preferred fit corresponds to $\alpha\simeq79^\circ$ in our convention, and an alternative treatment of a polarization jump also favoured a nearly orthogonal rotator.

Selection is a second difficulty. The Gould and Rankin radio samples contain many pulsars at intermediate inclinations. A narrow beam centred on an inclined magnetic axis, however, sweeps out a wider belt on the sky than the same beam close to alignment. When \citet{TaurisManchester1998} corrected for this effect with a circular dipolar beam, their inferred parent population favoured small inclinations. They also found evidence for alignment on a time-scale of order $10^7$ yr. The age dependence can be examined separately in the two catalogues, as \citet{Young2010} did after excluding low-field objects. A sample that mixes characteristic ages, $\tau_{\rm c}=P/(2\dot P)$, can thus hide a trend with inclination even within a single observing band.

\subsection{Inclination counts from radio and gamma-ray samples}

Figure~\ref{fig:obs_angle_counts} counts pulsars in $10^\circ$ intervals from published tables of preferred or best-fitting geometries. The FAST radio sample is concentrated at small and intermediate angles, with its largest bin at $20$--$30^\circ$ and few fits below $10^\circ$. The peak looks sharper than the measurements are: four in five of the tabulated fits have formal inclination intervals spanning more than $40^\circ$ after folding the two poles together. In the smaller radio study of young $\gamma$-ray pulsars, more than half of the preferred inclinations lie below $40^\circ$ \citep{Rookyard2015,RookyardInclination2015}. Random axes would put a fraction $1-\cos40^\circ$, less than a quarter, of a sample there before selection. At the other end of this histogram, half of the entries in the last bin were assigned $90^\circ$ by the original authors.

Twenty-four pulsars occur in both panels (b) and (c), which allows a direct comparison. Keeping the stars and the $40^\circ$ cut fixed, 14 lie below the cut according to their radio geometries, two according to the slot-gap fits and none according to the outer-gap fits. All three estimates refer to the same magnetic axes. A shift of this size need not reflect a change in the stars or in the site from which they radiate; it can arise from the way a pulse profile is converted into an angle. Millisecond pulsars show a related ambiguity. For the same 40 objects in panel (d), twice as many have $\alpha\geq60^\circ$ in the outer-gap family as in the two-pole caustic family.

\subsection{What the samples can test}

At present the inclination histograms do not resolve a signature of trapping. The radio fits favour smaller angles in several samples, but for the same stars the gamma-ray fits can shift the distribution by more than the width of the low-angle shoulder. The comparison with the score of Section~\ref{sec:emission_score} shows the same ambiguity. Taken at face value, the light-curve fits place nearly all of the 24 common pulsars above the $40^\circ$ cut, towards the range the score favours, while the radio geometries place 14 of them below it. The outcome depends on which set of angles is used, so neither tests the model.

There is also a mismatch in what is counted. The histogram counts each detected star once; the fraction in Eq.~\eqref{eq:footpoint_fraction} counts possible paths within a star of specified inclination. Even an exact angle for every object would leave the relation between these two counts undetermined.

Confinement could reduce the flux feeding an outer accelerator if particles are injected below the relevant barrier or lose energy along the way. Emission from charges retained at smaller radii would be a different channel, whose strength cannot be inferred from the number of wells. A useful test would compare an identifiable outer pulse or spectral component at similar spin-down power and age, using independent inclination estimates, distances, beam coverage and upper limits from a common survey. A quantitative prediction would also require the injected energy distribution and a field geometry appropriate to the emitting region.

%% file: 09_conclusions.tex
\section{Conclusions}
\label{sec:conclusions}

We have studied a bead that slides without friction on a field line of a rigidly rotating dipole, for arbitrary magnetic inclination and azimuth. The model is idealized: the shape of the line is prescribed, and there is no parallel electric field, gravity or radiation reaction. Our main results are as follows.

\begin{enumerate}
\item  The centrifugal force makes every local maximum of the distance from the rotation axis a stable equilibrium, provided the point lies outside the star and inside the light cylinder. The equilibrium angles are the roots of a cubic in $\tan\theta$ and depend only on the inclination and the magnetic azimuth; the size of the line decides whether a given root lies between the two boundaries. A tilted line has either one maximum, or two maxima separated by a minimum. Without drag, a bead released inside a well oscillates with constant amplitude.

\item  Two lines with the same potential can behave differently at the cylinder, because the outcome there is decided by the bend of the line. On a line that leads the rotation the bead approaches the cylinder exponentially in laboratory time and never arrives, although its proper time stays finite. On the mirror-image trailing line it crosses in a finite time, at a speed below that of light. A line in the tilt plane is the boundary case: the solution turns back at the point where the bead would have to move at the speed of light.

\item  The Jacobi integral gives a simple exclusion. A bead with $\mathcal E>1$ has no turning point inside the cylinder and passes every well. Particles injected relativistically along the field deep inside the cylinder are far above this threshold. The wells can hold only particles that are created slowly inside them or that have lost energy, and a charge that merely corotates in a well radiates very little.

\item  In the reference sample the mean number of wells per line roughly doubles between alignment and orthogonality. The increase survives other ranges and weightings of the lines, with a different magnitude. Wells seldom lie on the route to the cylinder, however: at $45^\circ$ only about one in fifteen paths that reach the cylinder meets a well first, and none does at alignment or orthogonality. On much longer lines the wells are confined to small inclinations and lie at spherical distances where a static dipole is a poor approximation.

\item  A simple score for emission from the outer magnetosphere is largest at inclinations between about $60^\circ$ and $75^\circ$, depending on the assumed yield. This range mainly reflects how many of the sampled lines reach the cylinder, and it moves to orthogonality for a different range of lines. Published radio and gamma-ray geometries for the same pulsars give inclination distributions that differ by more than the effect we look for. Present measurements therefore neither support nor exclude an inclination that favours trapping or outer emission.
\end{enumerate}

A natural next step is to replace the static dipole by a retarded or force-free field. There the open lines bend back near the light cylinder, so the finite-time crossing found here on trailing lines may be the relevant case. Injection and losses should also be included, since they decide whether the wells are populated at all.

%% file: appendices.tex
\section{Dipole coefficients for the laboratory equation}
\label{app:coeffs}
For direct numerical use of Eq.~\eqref{eq:general_theta_lab}, write
\begin{equation}
s=\sin\theta,\qquad c_\theta=\cos\theta,\qquad
W=1+3c_\theta^2,
\end{equation}
and define
\begin{align}
\kappa={}&\Omega\sin\alpha\sin\phi_m,\\
Q={}&\sin^2\alpha\,c_\theta^2
+2\sin\alpha\cos\alpha\cos\phi_m\,s c_\theta
\nonumber\\
&+\bigl(\cos^2\alpha\cos^2\phi_m+\sin^2\phi_m\bigr)s^2,\\
J={}&2\sin^2\alpha\,c_\theta^3
+5\sin\alpha\cos\alpha\cos\phi_m\,s c_\theta^2
\nonumber\\
&+\Bigl[3\bigl(\cos^2\alpha\cos^2\phi_m+\sin^2\phi_m\bigr)
-\sin^2\alpha\Bigr]s^2c_\theta
\nonumber\\
&-\sin\alpha\cos\alpha\cos\phi_m\,s^3.
\label{eq:dipole_eom_J}
\end{align}
In this notation the required coefficients and derivatives are
\begin{align}
A&=R^2s^2W, & A'&=4R^2s c_\theta(3c_\theta^2-1),\\
B&=R^2\kappa s^4, & B'&=4R^2\kappa s^3c_\theta,\\
C&=\Omega^2R^2s^4Q, & C'&=2\Omega^2R^2s^3J.
\end{align}
With
\begin{equation}
D=1-\Omega^2R^2s^4Q,\qquad
\Lambda=WD+R^2\kappa^2s^6,
\end{equation}
we have $K=R^2s^2\Lambda$. The explicit general equation is
\begin{equation}
\frac{d^2\theta}{dt^2}
=\frac{N_0+N_1v+N_2v^2+N_3v^3}{\Lambda},
\quad v=\frac{d\theta}{dt},
\label{eq:explicit_dipole_eom}
\end{equation}
where every coefficient is an explicit function of $\theta$:
\begin{align}
N_0={}&\Omega^2D sJ,\\
N_1={}&-3\Omega^2R^2\kappa s^5J,\\
N_2={}&-2\Omega^2R^2W s^3J
-\frac{2D c_\theta(3c_\theta^2-1)}{s}
\nonumber\\
&-4R^2\kappa^2s^5c_\theta,\\
N_3={}&-6R^2\kappa s^3c_\theta(1+c_\theta^2).
\label{eq:explicit_dipole_coefficients}
\end{align}
No restriction to a symmetry plane or a particular tilt has been
made. The apparent $1/s$ factor reflects the degeneracy of the
$\theta$ coordinate at the ideal dipole's endpoints; the equations
above apply to regular interior points $0<\theta<\pi$.

\section{The straight rotating line}\label{app:straight}

To recover the straight-channel model of \citet{MachabeliRogava1994}, consider
an orthogonal rotator with $\phi_m=0$ and take
$R\to\infty$, $\theta\to0$, keeping $r=R\theta^2$ finite.
The trajectory approaches
\begin{equation}
x=r\cos(\Omega t),\qquad
y=r\sin(\Omega t),\qquad z=0.
\end{equation}
The induced metric becomes
\begin{equation}
ds^2=-(1-\Omega^2r^2)\,dt^2+dr^2.
\end{equation}

Writing $v=dr/dt$, the conserved energy is
\begin{equation}
\mathcal E=
\frac{1-\Omega^2r^2}
{\sqrt{1-\Omega^2r^2-v^2}}.
\end{equation}
Differentiating this expression gives the laboratory-time
equation of motion
\begin{equation}
\frac{d^2r}{dt^2}
=
\frac{\Omega^2r}{1-\Omega^2r^2}
\left[
1-\Omega^2r^2
-2\left(\frac{dr}{dt}\right)^2
\right].
\end{equation}

\section{The aligned dipole in flat spacetime}
\label{app:aligned_flat}

When the magnetic and rotation axes coincide ($\alpha=0$), the mixed
coefficient $B$ in Section~\ref{sec:model} vanishes. On either side
of the magnetic equator, we can use the spherical radius
$r=R\sin^2\theta$ as the coordinate along the line, that is $\lambda=r$ in the notation of Section~\ref{sec:general}. In this coordinate,
the general first integral~\eqref{eq:equilibrium_energy_integral}
reduces to the radial velocity law~\eqref{eq:aligned_flat_radial_velocity}
derived below.

In spherical coordinates the Minkowski line element is
\begin{equation}
    ds^2 = -dt^2 + dr^2 + r^2 d\theta^2 + r^2 \sin^2\theta\,d\phi^2 .
\end{equation}
Along the dipole line, $\sin^2\theta=r/R$ and
$d\theta^2=dr^2/[4r(R-r)]$. Corotation gives $d\phi=\Omega\,dt$,
so the induced line element becomes
\begin{equation}
    ds^2 = -\left(1 - \frac{r^3 \Omega^2}{R}\right) dt^2
           + \left(1 + \frac{r}{4(R - r)}\right) dr^2 .
\end{equation}
For a bead of unit mass, the Lagrangian is
\begin{equation}
    \mathcal{L} = -\frac{1}{2}\left(1 - \frac{r^3 \Omega^2}{R}\right)\dot{t}^{\,2}
                  + \frac{1}{2}\left(1 + \frac{r}{4(R - r)}\right)\dot{r}^{\,2},
\end{equation}
where dots denote derivatives with respect to proper time $\tau$.
Since $t$ is a cyclic coordinate, its conjugate momentum is conserved.
Writing $E=\mathcal E$ for the Jacobi integral per unit rest mass, we have
\begin{equation}
    \frac{\partial \mathcal{L}}{\partial \dot{t}}
    = -\left(1 - \frac{r^3 \Omega^2}{R}\right)\dot{t} = -E ,
\end{equation}
and hence
\begin{equation}
    \frac{dt}{d\tau} = \frac{E}{1 - \frac{r^3 \Omega^2}{R}} .
\end{equation}
The four-velocity normalization $g_{\mu\nu}u^\mu u^\nu=-1$ gives
\begin{equation}
    -1 = -\left(1 - \frac{r^3 \Omega^2}{R}\right)\dot{t}^{\,2}
         + \left(1 + \frac{r}{4(R - r)}\right)\dot{r}^{\,2}.
\end{equation}
Dividing by $\dot{t}^{\,2}$ and writing $v=dr/dt$, we obtain
\begin{equation}
    -\left(\frac{d\tau}{dt}\right)^2
    = -\left(1 - \frac{r^3 \Omega^2}{R}\right)
      + \left(1 + \frac{r}{4(R - r)}\right)v^2 .
\end{equation}
Substituting $d\tau/dt=(1-r^3\Omega^2/R)/E$ gives the radial
velocity on the outward branch,
\begin{equation}
    v(r) = \sqrt{\frac{1 - \frac{r^3 \Omega^2}{R}}{1 + \frac{r}{4(R - r)}}
                 \left(1 - \frac{1 - \frac{r^3 \Omega^2}{R}}{E^2}\right)} .
    \label{eq:aligned_flat_radial_velocity}
\end{equation}

\section{The aligned dipole in Schwarzschild spacetime}
\label{app:aligned_schwarzschild}

We keep the same prescribed dipole line and include the gravity of a
central mass $M$. This is an example of the general treatment of Section~\ref{sec:general} with $g_{t\phi}=0$, rigid rotation and $\lambda=r$. In units $G=c=1$, the Schwarzschild line element is
\begin{equation}
    \begin{aligned}
    ds^2 ={}& -\left(1 - \frac{2M}{r}\right) dt^2
              + \left(1 - \frac{2M}{r}\right)^{-1} dr^2 \\
            & + r^2 d\theta^2 + r^2 \sin^2\theta\,d\phi^2 .
    \end{aligned}
\end{equation}
Imposing the dipole and corotation constraints of
Appendix~\ref{app:aligned_flat} gives the induced line element
\begin{equation}
    \begin{aligned}
    ds^2 ={}& -\left(1 - \frac{2M}{r} - \frac{r^3 \Omega^2}{R}\right) dt^2 \\
            & + \left(\frac{r}{r - 2M} + \frac{r}{4(R - r)}\right) dr^2 .
    \end{aligned}
\end{equation}
As in flat spacetime, $t$ is cyclic. The conserved energy per unit
rest mass $E$ therefore gives
\begin{equation}
    \frac{dt}{d\tau} = \frac{E}{1 - \frac{2M}{r} - \frac{r^3 \Omega^2}{R}} .
\end{equation}
Using the components of the induced metric, the four-velocity
normalization reads $-(d\tau/dt)^2=g_{tt}+g_{rr}v^2$.
The outward radial velocity in Schwarzschild coordinate time is then
\begin{equation}
    v(r) = \sqrt{\frac{1 - \frac{2M}{r} - \frac{r^3 \Omega^2}{R}}
                      {\frac{r}{r - 2M} + \frac{r}{4(R - r)}}
                 \left(1 - \frac{1 - \frac{2M}{r} - \frac{r^3 \Omega^2}{R}}{E^2}\right)} .
\end{equation}
Setting $M=0$ recovers Eq.~\eqref{eq:aligned_flat_radial_velocity}.